\documentclass[lettersize, journal]{IEEEtran}
\usepackage[T1]{fontenc}
\usepackage{cite}
\usepackage{url}
\usepackage{amsmath,amssymb,amsfonts}
\usepackage{graphicx}
\usepackage{textcomp}
\usepackage[table]{xcolor}
\usepackage{bm}
\usepackage{amsthm}
\usepackage{enumerate}
\usepackage{booktabs}
\usepackage{diagbox}
\usepackage{supertabular}
\usepackage{subfigure}
\usepackage{caption}
\usepackage{algorithmicx}
\usepackage{comment}
\usepackage{multirow}
\usepackage{filecontents} 
\usepackage{flushend}
\usepackage{algorithmicx}
\usepackage[linesnumbered,ruled,vlined]{algorithm2e}
\newtheorem{definition}{Definition}
\newtheorem{theorem}{Theorem}
\newtheorem{lemma}{Lemma}

\newtheorem{remark}{Remark}

\definecolor{seagreen}{rgb}{0.18, 0.55, 0.34}
\definecolor{royalpurple}{rgb}{0.47,0.32,0.66}
\definecolor{brown(traditional)}{rgb}{0.59, 0.29, 0.0}
\definecolor{blue}{rgb}{0.3, 0.2, 0.9}
\usepackage[colorlinks,
            linkcolor=blue,
            anchorcolor=blue,
            citecolor=blue]{hyperref}

\begin{document} 
\title{
Diffusion-based Dynamic Contract for Federated AI Agent Construction in Mobile Metaverses
}

\author{
Jinbo Wen, Jiawen Kang, Yang Zhang, Yue Zhong, Dusit Niyato, \textit{Fellow, IEEE}, Jie Xu, \textit{Fellow, IEEE}, \\ Jianhang Tang, \textit{Member, IEEE}, and Chau Yuen, \textit{Fellow, IEEE}
\thanks{
J. Wen and Y. Zhang are with the College of Computer Science and Technology, Nanjing University of Aeronautics and Astronautics, China (e-mails: jinbo1608@nuaa.edu.cn; yangzhang@nuaa.edu.cn). 
J. Kang and Y. Zhong are with the School of Automation, Guangdong University of Technology, China (e-mails: kavinkang@gdut.edu.cn; 2112404106@mail2.gdut.edu.cn). 
D. Niyato is with the College of Computing and Data Science, Nanyang Technological University, Singapore (e-mail: dniyato@ntu.edu.sg). 
J. Xu is with the School of Science and Engineering (SSE), the Shenzhen Future Network of Intelligence Institute (FNii-Shenzhen), and the Guangdong Provincial Key Laboratory of Future Networks of Intelligence, The Chinese University of Hong Kong (Shenzhen), China (e-mail: xujie@cuhk.edu.cn). 
J. Tang is with the State Key Laboratory of Public Big Data, Guizhou University, China (e-mail: jhtang@gzu.edu.cn). 
C. Yuen is with the School of Electrical and Electronics Engineering, Nanyang Technological University, Singapore (e-mail: chau.yuen@ntu.edu.sg).
}}

\markboth{IEEE TRANSACTIONS ON SERVICES COMPUTING}{Diffusion-based Dynamic Contract for Federated AI Agent Construction in Mobile Metaverses}

\maketitle

\begin{abstract}
Mobile metaverses are envisioned as a transformative digital ecosystem that delivers immersive, intelligent, and ubiquitous services through mobile devices. Driven by Large Language Models (LLMs) and Vision-Language Models (VLMs), Artificial Intelligence (AI) agents hold the potential to empower the creation, maintenance, and evolution of mobile metaverses, enabling seamless human-machine interaction and dynamic service adaptation. Currently, AI agents are primarily built upon cloud-based LLMs and VLMs. However, several challenges hinder their efficient deployment, including high service latency and a risk of sensitive data leakage during perception and processing. In this paper, we develop an edge-cloud collaboration-based federated AI agent construction framework in mobile metaverses. Specifically, Edge Servers (ESs), as agent infrastructures, first create agent modules in a distributed manner. The cloud server then integrates these modules into AI agents and deploys them at the edge, thereby enabling low-latency AI agent services for users. Considering that ESs may exhibit dynamic levels of willingness to participate in federated AI agent construction, we design a two-period dynamic contract model to continuously incentivize ESs to participate in agent module creation, effectively addressing the dynamic information asymmetry between the cloud server and ESs. Furthermore, we propose an Enhanced Diffusion Model-based Soft Actor-Critic (EDMSAC) algorithm to effectively generate optimal dynamic contracts. In the algorithm, we apply dynamic structured pruning to DM-based actor networks to enhance denoising efficiency and policy learning performance. Simulation results demonstrate that the EDMSAC algorithm outperforms the DMSAC algorithm by up to $23\%$ in optimal dynamic contract generation.

\end{abstract}


\section{Introduction}
\IEEEPARstart{W}{ith} the advancement of extended reality, Generative Artificial Intelligence (GenAI), and wireless communication technologies, mobile metaverses are emerging as a transformative digital ecosystem~\cite{10472663}. Positioned as a cornerstone of the next-generation Internet, mobile metaverses have attracted significant attention from both academia and industry due to their robust levels of connectivity, enabling users to interact in real time with immersive and interconnected virtual spaces through mobile devices~\cite{9944868}. Furthermore, mobile metaverses hold the potential to revolutionize various fields, such as healthcare, education, and entertainment~\cite{9944868}. However, the development of mobile metaverses faces several technical challenges. Firstly, conventional content creation methods often involve time-consuming and resource-intensive processes, leading to static virtual content that fails to dynamically adapt to real-time user interactions or environmental variations~\cite{10841385}. Secondly, the inherent limitations of mobile input modalities, such as touchscreen gestures and voice commands, would inevitably degrade the fluidity of immersive interactions.

GenAI is driving the evolution of the metaverse due to its incredible ability to generate novel content~\cite{10517486}. As an advanced paradigm of GenAI, AI agents are expected to play a pivotal role in the creation, maintenance, and evolution of mobile metaverses. Empowered by Large Language Models (LLMs) and Vision-Language Models (VLMs)~\cite{durante2024agentaisurveyinghorizons}, AI agents are artificial entities comprising four core modules: perception, memory, decision-making, and action. With these modules, AI agents are capable of perceiving surrounding environments, processing and storing information, formulating strategies, and executing tasks to achieve specific objectives~\cite{zhang2024proagent}. Within mobile metaverses, AI agents can improve immersion by dynamically generating and adapting virtual content, as well as creating realistic avatars and environments using Digital Twin (DT) and three-dimensional rendering technologies~\cite{durante2024agentaisurveyinghorizons}. Moreover, they can deliver personalized and interactive user experiences by acting as intelligent and personalized assistants, guiding users through virtual environments, and tailoring interactions based on individual preferences.

Currently, AI agent construction often relies on cloud-based LLMs and VLMs, which require transmitting perceived data to remote servers for processing before returning results to users~\cite{durante2024agentaisurveyinghorizons, zhang2024proagent}. Although AI agents hold significant promise for enabling mobile metaverses, their efficient implementation is hindered by several critical challenges: 1) The cloud-based AI agent architecture introduces inherent service latency, negatively impacting user experience. For example, latency in processing and rendering real-time location data within augmented reality navigation systems may lead to user disorientation, operational inefficiencies, or even safety hazards in scenarios requiring precise spatial awareness~\cite{siriwardhana2021survey}. 2) During peak usage periods, cloud servers may experience performance bottlenecks due to simultaneous service requests from large user bases~\cite{10847942}. Such bottlenecks result in response latency and potential service unavailability under high-load conditions, undermining the seamless interaction demanded by immersive metaverse environments. 3) The transmission of perceived data, which often contains sensitive personal information, remains vulnerable to interception during transmission and susceptible to breaches if stored centrally in cloud repositories~\cite{nguyen2021security}. This exposes users to risks such as unauthorized tracking, data misuse, and identity theft, particularly in mobile metaverses where data flows across heterogeneous networks.

Motivated by the feasibility of federated model construction~\cite{10254627, sun2020adaptive, yang2022optimizing, 10679522, 10443270}, we propose a novel edge-cloud collaboration-based federated AI agent construction framework, where Edge Servers (ESs) with sufficient computing and storage resources serve as agent infrastructures responsible for constructing agent modules in a distributed manner, while the cloud server is responsible for integrating these constructed modules into AI agents and eventually deploys them at the edge. Unlike traditional federated learning, which deals with the same task across participating nodes~\cite{9264742}, the agent modules in the federated AI agent construction framework can be heterogeneous, with each module performing distinct functions. Furthermore, whereas federated learning typically aims to jointly train a single global model until convergence~\cite{9264742}, the construction of multiple agent modules can be done concurrently. Nonetheless, during the construction of agent modules, ESs exhibit different dynamic levels of Willingness to Participate (WTP) over multiple periods due to various factors~\cite{DynamicLim}, such as the high energy consumption of agent module construction and the prioritization of other computing tasks. Moreover, the WTP levels of ESs are typically unknown to the cloud server, leading to the intention of some ESs to free-ride on the efforts of others~\cite{DynamicLim}. To this end, we design a dynamic contract-based incentive mechanism to incentivize ESs to participate sustainably in federated AI agent construction, addressing the challenge of \textit{dynamic information asymmetry}.

Since dynamic contract design in practical scenarios typically involves changing network conditions~\cite{DynamicLim}, conventional optimization techniques that rely on complete prior knowledge of the environment must be frequently redesigned~\cite{wen2024sustainablediffusionbasedincentivemechanism}, which may not be applicable in practical scenarios. Fortunately, Deep Reinforcement Learning (DRL) algorithms have been widely utilized in dynamic scenarios, which can learn optimal strategies through repeated interactions with the environment~\cite{10368063}. Moreover, Generative Diffusion Models (GDMs) with a robust generative capability have shown a strong potential to improve DRL performance~\cite{10529221, 10409284, wang2023diffusionpoliciesexpressivepolicy}, particularly by mitigating the risk of convergence to suboptimal solutions. Therefore, we adopt an enhanced diffusion-based DRL algorithm to identify optimal dynamic contracts. The main contributions of this paper are summarized as follows:
\begin{itemize}
    \item \textbf{Federated AI Agent Construction Framework:} We propose an edge-cloud collaboration-based federated AI agent construction framework in mobile metaverses. In the proposed framework, ESs serve as agent infrastructures, leveraging their computing, storage, and communication resources to construct agent modules in a distributed manner. The cloud server then integrates these created agent modules into complete AI agents and deploys them at the edge, thereby providing low-latency AI agent services for users in mobile metaverses.
    \item \textbf{Dynamic Contract-based Incentive Mechanism:} Considering the dynamic WTP levels of ESs in federated AI agent construction, we propose a two-period dynamic contract model to motivate ESs to continuously participate in agent module creation. The self-disclosure property of the proposed model ensures that each ES selects the contract best suited to its WTP level, thereby addressing the problem of dynamic information asymmetry. Notably, the two-period dynamic contract design maintains relatively low computational complexity and thus can be recursively extended to multiple periods.
    \item \textbf{Diffusion Models for Dynamic Contract Design:} We adopt an Enhanced DM-based Soft Actor-Critic (EDMSAC) algorithm to identify optimal dynamic contracts. Specifically, DM-based actor networks possess both analytical and generative capabilities, which can generate dynamic contract samples through an iterative denoising process. To enhance denoising efficiency, we incorporate dynamic structured pruning techniques into DM-based actor networks, enabling a more efficient and scalable exploration of the optimal dynamic contract policy. \textit{To the best of our knowledge, this is the first work to leverage DRL algorithms, especially diffusion-based DRL, for optimal dynamic contract design}.
    \item \textbf{Extensive Simulations for Performance Evaluation:} We utilize LLMs called Qwen-VL\footnote{\url{https://github.com/QwenLM/Qwen-VL?tab=readme-ov-file}} to construct an AI agent capable of operating mobile phones to perform simple tasks. Specifically, we leverage Qwen-VL-Plus to create a perception module and Qwen-VL-Max to create decision-making, reflection, and planning modules. This construction process can be extended to the proposed framework. We then compare the EDMSAC algorithm with two DM-based DRL algorithms and three representative DRL benchmark algorithms. Simulation results demonstrate that the EDMSAC algorithm can achieve up to a $23\%$ improvement over the DMSAC algorithm~\cite{10409284} in optimal dynamic contract generation.
\end{itemize}

The remainder of the paper is organized as follows: Section \ref{Related} reviews the related work, where Table \ref{LiteratureComparison} summarizes the comparison of the current literature and this paper in optimal contract design. Section \ref{EC_Framework} introduces the edge-cloud collaboration-based federated AI agent construction framework. In Section \ref{Dynamic_Contract_Design}, we propose the two-period dynamic contract model to motivate ESs to continuously participate in federated AI agent construction. Section \ref{EDMSAC} presents the EDMSAC algorithm for generating optimal dynamic contracts. In Section \ref{Simulation}, we provide extensive simulation results to demonstrate the feasibility and effectiveness of the proposed framework and algorithm. Section \ref{Conclusion} concludes the paper.

\begin{table*}[t]
\centering
\caption{Comparison Between the Current Literature and This Paper in Optimal Contract Design}
\label{LiteratureComparison}
\renewcommand{\arraystretch}{1.3}
\setlength{\tabcolsep}{5pt}
\begin{tabular}{c|c|ccccccccccccc}
\toprule
\rowcolor{gray!6}
\multicolumn{2}{c|}{\textbf{Literature}} 
& \cite{10254627} & \cite{DynamicLim} & \cite{wen2024sustainablediffusionbasedincentivemechanism} & \cite{10368063} & \cite{JinboDiffIoT} & \cite{10570575} & \cite{9440722} & \cite{DCTITS} & \cite{9106861} & \cite{10829636} & \cite{wen2024learningbasedbigdatasharing} & \cite{10288549} & \textbf{Our Paper} \\
\midrule
\multirow{3}{*}{\textbf{Properties}} 
& Individual Rationality 
& \checkmark & \checkmark & \checkmark & \checkmark & \checkmark & \checkmark 
& \checkmark & \checkmark & \checkmark 
& \checkmark & \checkmark & \checkmark & \textcolor{red}{\checkmark} \\
& Incentive Compatibility    
& \checkmark & \checkmark & \checkmark & \checkmark & \checkmark & \checkmark 
& \checkmark & \checkmark & \checkmark 
& \checkmark & \checkmark & \checkmark & \textcolor{red}{\checkmark} \\
& Intertemporal Consistency    
& & \checkmark & & & &  
& & \checkmark & & & & & \textcolor{red}{\checkmark} \\
\midrule
\multirow{3}{*}{\textbf{Solutions}} 
& Deep Reinforcement Learning 
&  &  & \checkmark & \checkmark & \checkmark& 
& & &  
& \checkmark & \checkmark & \checkmark & \textcolor{red}{\checkmark} \\
& Diffusion Models    
& & & \checkmark & & \checkmark& 
& & & 
&  \checkmark &  & & \textcolor{red}{\checkmark} \\
& Enhanced Diffusion Models    
& & & \checkmark & & &  
& & & & & & & \textcolor{red}{\checkmark} \\
\bottomrule
\end{tabular}
\end{table*}

\section{Related Work}\label{Related}

\subsection{Federated Model Construction}
Federated Learning (FL) is a distributed Machine Learning (ML) framework that enables collaborative model training while preserving data privacy through secure aggregation, making it particularly suitable for applications with stringent privacy requirements~\cite{10896832}. Recent research has highlighted the effectiveness and potential of FL in constructing DT and ML models, particularly in enhancing privacy protection and optimizing resource utilization~\cite{10254627, sun2020adaptive, yang2022optimizing, 10679522, 10443270}. For instance, the authors in~\cite{sun2020adaptive} combined FL with DT technology and introduced a trust mechanism to optimize the learning process. 
Similarly, the authors in~\cite{yang2022optimizing} leveraged FL to optimize DT-enabled industrial Internet of Things systems. 
In~\cite{10679522}, the authors proposed an FL architecture based on over-the-air computing, enabling the efficient and secure development of DT models. 
In~\cite{10443270}, the authors proposed a generalized federated Reinforcement Learning (RL) framework that incorporates meta-learning techniques, which enables the fusion of RL models trained across multiple intelligent devices into a unified and generalizable model. 

Combined with a device-edge-cloud collaborative computing architecture, FL can facilitate efficient model training by enabling small-scale learning on devices while leveraging the cloud for large-scale model updates. This decentralized approach minimizes network dependency, significantly enhancing the responsiveness and operational stability of AI agents and making it a promising solution for AI agent model construction. To the best of our knowledge, there is currently no existing study on the construction of AI agents in wireless mobile environments. Motivated by the aforementioned studies, we aim to develop a novel federated AI agent construction framework, exploring the feasibility and potential of mobile AI agents in wireless networks.

\subsection{Contract Theory-based Wireless Applications}
Contract theory is an economic tool that examines cooperation in settings with incomplete information and designs incentives for agents to participate in tasks involving asymmetric information~\cite{JinboDiffIoT}. In the context of FL, the authors in~\cite{10570575} proposed an incentive contract tailored for scenarios with strong information asymmetry, leveraging an iteration algorithm to determine optimal contract elements. In~\cite{9440722}, the authors utilized contract theory to motivate privately owned vehicles to contribute their onboard computing resources. 
Traditional contracts are generally one-dimensional, accounting for a static type of information asymmetry between the two parties involved~\cite{DCTITS}. Therefore, there is a growing interest in multi-dimensional and dynamic contracts that account for multiple asymmetries and evolving conditions. 

As for the applications of multi-dimensional contracts~\cite{9106861,10829636}, the authors in~\cite{9106861} developed a multi-dimensional contract model to optimize the data rewarding design in mobile networks under asymmetric information. In~\cite{10829636}, the authors constructed a multi-dimensional contract theory-based framework, incorporating Prospect Theory (PT) and GDM-based DRL algorithms to optimize the dynamic migration of embodied AI twins in vehicular embodied AI networks. As for the applications of dynamic contracts~\cite{DCTITS, DynamicLim}, the authors in~\cite{DynamicLim} proposed a two-period contract-based incentive mechanism to motivate users to participate in FL training under dynamic information asymmetry. In~\cite{DCTITS}, 
the authors proposed a dynamic contract-based incentive mechanism to motivate vehicles to share their threat data for LLM fine-tuning. Inspired by the above works, we propose a dynamic contract-based incentive mechanism to motivate ESs to continuously participate in AI agent construction.

\subsection{Deep Reinforcement Learning for Contract Design}
RL algorithms have been extensively used to identify optimization strategies~\cite{RAHMANI2024109514, RAHMANI2025103804}, including optimal static contract design~\cite{wen2024learningbasedbigdatasharing, 10288549, 10368063}. For instance, the authors in~\cite{wen2024learningbasedbigdatasharing} employed Proximal Policy Optimization (PPO) algorithms to identify optimal contracts, incentivizing mobile devices to share data in mobile AI-Generated Content (AIGC) networks. In~\cite{10288549}, the authors utilized the twin-delayed Deep Deterministic Policy Gradient (DDPG) algorithm to determine the incentive reward policy. Nevertheless, traditional DRL algorithms often struggle to directly identify optimal contracts due to the complex Individual Rationality (IR) and Incentive Compatibility (IC) constraints. Fortunately, GDMs excel at capturing high-dimensional information in dynamic and complicated network optimization scenarios~\cite{10529221}. Several studies have explored the use of GDMs for optimal static contract design~\cite{JinboDiffIoT, 10829636, wen2024sustainablediffusionbasedincentivemechanism}. For example, the authors in~\cite{JinboDiffIoT} proposed a GDM-based DRL algorithm to generate optimal contracts under PT, addressing the problem of information asymmetry in edge AIGC service provision. However, existing studies still rely on heuristic algorithms for optimal dynamic contract design~\cite{DynamicLim, DCTITS}, which may not be practical in real scenarios. To fill this gap, we propose the EDMSAC algorithm to identify optimal dynamic contracts. 

\section{Edge-Cloud Collaboration-based Federated AI Agent Construction Framework}\label{EC_Framework}
In this section, we first introduce the proposed framework for constructing AI agents through edge-cloud collaboration. Then, we present the model quality and energy consumption for federated AI agent construction. Finally, we formulate the utilities of ESs and the cloud server, respectively. 

\subsection{Framework Design}~\label{framework_design}
For federated AI agent construction, we consider a central cloud server and a set of geographically distributed ESs, which are usually installed in cellular base stations or roadside units~\cite{10829636}, denoted by $\mathcal{N} = \{1,\ldots, n, \ldots, N\}$. These ESs possess computing, storage, and communication capabilities sufficient to construct, cache, and transmit agent modules~\cite{chen2025federated}. During the AI agent construction period, AI agents can be split into multiple agent modules, such as perception, memory, and action~\cite{durante2024agentaisurveyinghorizons, zhang2024proagent}, each responsible for specific tasks based on the featured data perceived by ESs.

\begin{figure}
    \centering
    \includegraphics[width=0.38\textwidth]{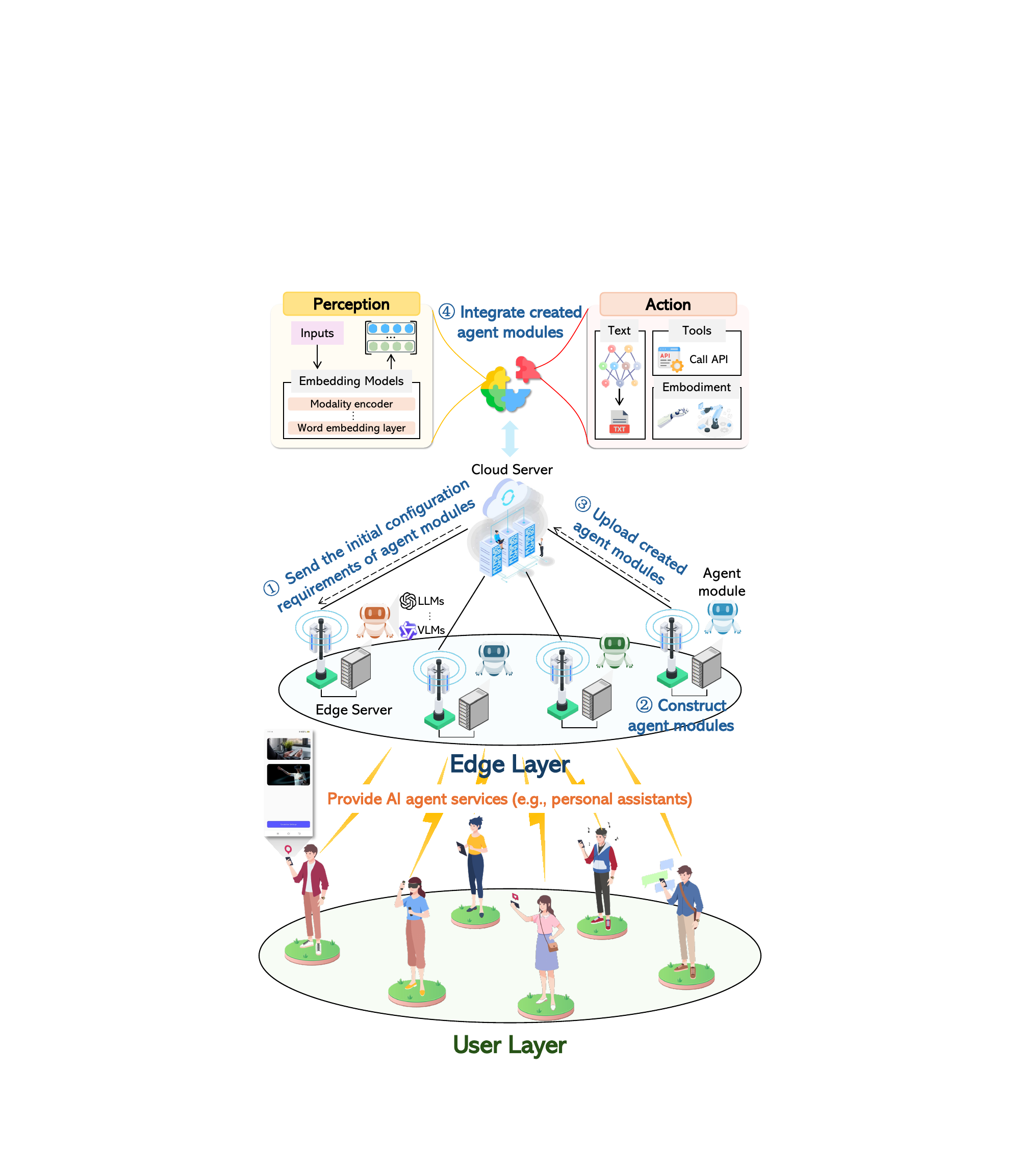}
    \caption{The proposed edge-cloud collaboration-based federated AI agent construction framework. Note that the interface of personal assistants is from the OmAgent application~\cite{zhang2024omagentmultimodalagentframework} downloaded by our Android phone.}
    \label{Framework}
\end{figure}

As shown in Fig. \ref{Framework}, before the beginning of agent module construction, the cloud server transmits the initial configuration requirements of agent modules to the ESs (Step \textcircled{1})~\cite{10254627}. The ESs, serving as agent infrastructures with sufficient resources such as computing, storage, and communication, leverage local LLMs and AI models (e.g., icon detection models) to construct the agent modules in a distributed manner (Step \textcircled{2})~\cite{wang2024mobileagentv2mobiledeviceoperation}. Upon completing the construction, the ESs upload the constructed agent modules to the cloud server (Step \textcircled{3})~\cite{10254627}. Finally, the cloud server creates the complete AI agents by integrating the uploaded agent modules, i.e., connecting and harmonizing these agent modules into a complete agent (Step \textcircled{4})~\cite{chen2025federated}. Once constructed, the AI agents can be deployed at the edge, allowing users to connect to them through the Internet Protocol (IP) address of the current local area network with the ESs~\cite{zhang2024omagentmultimodalagentframework}, facilitating low-latency AI agent services.

\subsection{Model Quality of Agent Modules}

Agent module construction involves fine-tuning pre-trained large models and other AI models to better process downstream tasks~\cite{durante2024agentaisurveyinghorizons}. For example, during the perception module construction, ESs are required to fine-tune visual perception models, such as vision transformer models, to extract corresponding semantic representations from visual inputs using samples of feature data. The extracted semantic content is then passed to LLMs for subsequent processing and reasoning. To evaluate the performance of agent module construction, we employ an $L$-Lipschitz continuous and $\varepsilon$-strongly convex loss function to approximately calculate the model quality of agent modules~\cite{chen2025federated, li2025collaborative}, which is a widely used definition of model accuracy in federated learning~\cite{10086671, chen2025federated}. We denote $Q_n$ as the model quality of agent modules created by ES $n$, which is calculated from the number of training rounds $T_n$ that ES $n$ determines to take for creating agent modules, as given by~\cite{10086671}
\begin{equation}\label{quality}
    Q_n = 1 - 2^\frac{-T_n(2-L\delta)\delta \varepsilon}{2},
\end{equation}
where $\delta \in (0, \frac{2}{L})$, $L$, and $\varepsilon$ are pre-defined hyperparameters~\cite{chen2025federated}. Given that the second derivative of $Q_n$ in (\ref{quality}) is calculated as $\frac{\partial^{2} Q_n}{\partial (T_n)^{2}} = -\big[\frac{(2-L\delta)\delta \varepsilon}{2} \ln2\big]^2 2^\frac{-T_n(2-L\delta)\delta \varepsilon}{2} < 0$, $Q_n$ is a convex function with respect to $T_n$, indicating that there exists the optimal number of training rounds for ESs to obtain the highest model quality of agent modules.

\subsection{Energy Consumption of Edge Servers}
We represent the total energy consumption of ES $n$ for federated AI agent construction as $E_n^{Tot}$, which consists of the energy consumption of perceiving feature data $E_n^{Per}$, creating agent modules $E_n^{Cre}$, and uploading created agent modules to the cloud server $E_n^{Upl}$. Specifically, the energy consumption of perceiving feature data $E_n^{Per}$ is given by~\cite{9729765}
\begin{equation}
    E_n^{Per} = \varsigma_n \Upsilon_n^{Per} D_n^{Fea} f_n^2,
\end{equation}
where $\varsigma_n$ is a constant related to the hardware architecture of ES $n$, $\Upsilon_n^{Per}$ ($\rm{cycles/bit}$) is the number of Central Processing Unit (CPU) cycles required for perceiving feature data, $D_n^{Fea}$ ($\rm{bits}$) is the size of feature data perceived by ES $n$, and $f_n$ ($\rm{Hz}$) is the CPU speed of ES $n$.

After perceiving feature data, ES $n$ constructs agent modules under iterative model training to improve the model quality of agent modules, and its energy consumption of creating agent modules $E_n^{Cre}$ is given by~\cite{9264742}
\begin{equation}
    E_n^{Cre} = \rho_n T_n \Upsilon_n^{Cre} f_n^2,
\end{equation}
where $\rho_n$ represents the effective switched capacitance of ES $n$ and $\Upsilon_n^{Cre}$ ($\rm{cycles/bit}$) is the number of CPU cycles required for creating one bit of agent modules on ES $n$.

Then, each ES $n$ uploads its constructed agent modules to the cloud server for integration, and the corresponding energy consumption $E_n^{Upl}$ is given by~\cite{9264742}
\begin{equation}
    E_n^{Upl} = \frac{P_n D_n^{Agent}}{r_{n, c}},
\end{equation}
where $P_n$ ($\rm{dBm}$) is the transmit power of ES $n$, $D_n^{Agent}$ ($\rm{bits}$) is the model size of agent modules constructed by ES $n$, and $r_{n, c}$ ($\rm{Mbps}$) represents the transmission rate from ES $n$ to the cloud server through the fiber link.

Therefore, the total energy consumption of ES $n$ for federated AI agent construction $E_n^{Tot}$ is expressed as
\begin{equation}
    E_n^{Tot} = (\varsigma_n \Upsilon_n^{Per} D_n^{Fea} + \rho_n T_n \Upsilon_n^{Cre}) f_n^2 + \frac{P_n D_n^{Agent}}{r_{n, c}}.
\end{equation}

\subsection{Utilities of Edge Servers and the Cloud Server}
For ES $n$, we define the level of WTP in federated AI agent construction as $\theta_n$~\cite{DynamicLim}, which reflects the perception, communication, and computing capabilities of ES $n$~\cite{10368063, wen2024sustainablediffusionbasedincentivemechanism}. The utility of ES $n$ is formulated based on the reward $R_n$ received from the cloud server and the total energy consumption $E_n^{Tot}$ for federated AI agent construction~\cite{DynamicLim,10368063}, which is given by
\begin{equation}\label{ESUtility}
    u_n = \theta_n R_n - c T_n - E,
\end{equation}
where $c = \sigma \rho_n \Upsilon_n^{Cre} f_n^2$ and $E = \sigma(\varsigma_n \Upsilon_n^{Per} D_n^{Fea} f_n^2  + \frac{P_n D_n^{Agent}}{r_{n, c}})$. Here, $\sigma$ is the unit cost of energy consumption.

Similarly, the utility of the cloud server is formulated based on the total profit obtained from all ESs and the cost of energy consumption of global AI agent integration $E_{C}$, as given by
\begin{equation}\label{utilityCS}
    U_{CS} = \sum_{n=1}^N(\alpha Q_n-R_n) - \sigma E_{C},
\end{equation}
where $\alpha$ represents a scaling factor that affects the overall magnitude of economic benefits~\cite{DCTITS}, and $E_{C}$ is influenced by several factors~\cite{10086671}, including the latency of global AI agent integration, both CPU and Graphics Processing Unit (GPU) resources on the cloud, and the average number of instructions in integrating agent modules.

Due to high workload and energy consumption, ESs may not always be willing to participate in federated AI agent construction, indicating that ESs have different levels of WTP for federated AI agent construction~\cite{DynamicLim}. The WTP levels of ESs are generally unknown to the cloud server and may change over multiple training rounds~\cite{DynamicLim}, which results in dynamic information asymmetry between the cloud server and ESs. Hence, it is necessary to develop a dynamic incentive scheme for ESs to participate in federated AI agent construction.

\section{Dynamic Contract Design}\label{Dynamic_Contract_Design}
In this section, we propose a two-period dynamic contract model to motivate ESs to continuously participate in federated AI agent construction under dynamic information asymmetry. Note that the proposed dynamic contract model possesses the inherent correlation between two periods within the whole agent construction time and can be feasibly extended to accommodate multiple periods recursively~\cite{DynamicLim, DCTITS}. In practice, the two-period dynamic contract design is efficient to balance incentive effectiveness and computing complexity~\cite{DynamicLim}.

\subsection{Two-Period Dynamic Contract Formulation}
Due to information asymmetry, the cloud server only knows the historical distribution of WTP levels of ESs through previous participation records rather than the precise WTP level of each ES~\cite{DynamicLim, JinboDiffIoT}. Thus, the cloud server can use data mining tools to divide ESs into a set $\Theta = \{\theta_k|\theta_k\in \mathbb{N}_+, \:k \in \mathcal{K} = \{1,\ldots, K,\: K \leq N\}\}$ of $K$ types based on the historical distribution of WTP levels~\cite{DynamicLim, JinboDiffIoT}. Considering the certain correlations in WTP across periods~\cite{DCTITS, DynamicLim}, in the first period, the $k$-th type of ESs can be denoted as $\theta_k^1,\:k\in\mathcal{K}$, and in the second period, the $k$-th type of ESs is denoted as $\theta_j^2(\theta_k^1),\:j\in\mathcal{K}$, which is an expectation correlated to $\theta_k^1$. Note that these two types are drawn independently from $\Theta$. Without loss of generality, the types in the first and second periods are sorted in non-decreasing order as $\theta_1^1 \leq\cdots\leq \theta_k^1 \leq\cdots\leq \theta_K^1$ and $\theta_1^2(\theta_k^1) \leq\cdots\leq \theta_j^2(\theta_k^1) \leq\cdots\leq \theta_K^2(\theta_k^1)$, respectively~\cite{DynamicLim, DCTITS}. We define $p_k^t$ as the probability that ESs belong to type $\theta_k^t$ during period $t$, where $\sum_{k=1}^K p_k^t = 1,\: t\in\{1,2\}$.

To incentivize ESs to continuously participate in federated AI agent construction, the cloud server issues a two-period dynamic contract $\Omega (\theta_k^1, \theta_j^2(\theta_k^1))$ at the beginning of the first period before the agent module creation starts. The contract item consists of WTP along with the corresponding reward. In the first period, the contract item designed for type-$k$ ESs is represented as $\{T_k^1(\theta_k^1), R_k^1(\theta_k^1)\}$, and in the second period, the contract item designed for the same type-$k$ ESs is represented as $\{T_j^2(\theta_j^2(\theta_k^1)), R_j^2(\theta_j^2(\theta_k^1))\},\: k,j\in \mathcal{K}$. Thus, the formulation of the two-period dynamic contract $\Omega$ is expressed as
\begin{equation}
    \Omega = \{(T_k^1(\theta_k^1), R_k^1(\theta_k^1)), (T_j^2(\theta_j^2(\theta_k^1)), R_j^2(\theta_j^2(\theta_k^1))) \}.
\end{equation}
For conciseness, we denote $T_j^2(\theta_k^1)$ and $R_j^2(\theta_k^1)$ as $T_j^2(\theta_j^2(\theta_k^1))$ and $R_j^2(\theta_j^2(\theta_k^1))$, respectively. Based on (\ref{ESUtility}), the utility of ESs with type $\theta_k^1$ in period 1 is given by
\begin{equation}
     u_k^1(T_k^1) = \theta_k^1 R_k^1(\theta_k^1) - c T_k^1(\theta_k^1) - E.
\end{equation}

Considering that the WTP level $\theta_k^1$ of type-$k$ ESs in period 1 would affect the utility in period 2~\cite{DCTITS}, the expected utility of type-$k$ ESs in period 2 is given by
\begin{equation}\label{equation10}
    U_k = u_k^1(T_k^1) + \underbrace{\beta \sum_{j = 1}^K p_j^2(T_k^1)u_j^2(T_j^2(T_k^1))}_{\rm{Expected}\:\rm{discounted}\:\rm{utility}},
\end{equation}
where $\beta \in (0, 1]$ and $p_j^2(T_k^1)$ represent a discount factor for the utility and the probability that type-$k$ ESs belong to type $\theta_j^2$ during period 2, respectively. The second term in (\ref{equation10}) represents the expected discounted utility across all $K$ types that ESs are characterized with certain probabilities in period 2, which is designed to capture the long-term benefit of ESs participating in the federated AI agent construction. A larger $\beta$ indicates a greater valuation of the future utility relative to the current utility~\cite{DCTITS, DynamicLim}. Since WTP levels across two periods are positively correlated~\cite{DynamicLim}, we have $p_j^2(T_k^1) \neq p_j^2(T^1_{k^{\prime}})$ and $\sum_{j = 1}^K p_j^2(T_k^1)u_j^2(T_j^2(T_k^1)) \neq \sum_{j = 1}^K p_j^2(T^1_{k^{\prime}})u_j^2(T_j^2(T^1_{k^{\prime}})),\: k \neq k^{\prime}$, with the proof presented in Lemma \ref{lemma7} in this paper.

Based on (\ref{utilityCS}), the expected utility of the cloud server in period $t \in \{1,2\}$ is derived as~\cite{DCTITS, DynamicLim}
\begin{equation}\label{expected_utility}
    U_{CS}^t = \Bigg[\sum_{k=1}^K p_k^t N \Big(\alpha \Big(1 - 2^\frac{-T_k^t(2-L\delta)\delta \varepsilon}{2}\Big)-R_k^t\Big)\Bigg]-\sigma E_{C}.
\end{equation}

Therefore, the total profit of the cloud server over two periods is expressed as
\begin{equation}\label{total_profit}
    U(\Omega) = U_{CS}^1 + \beta U_{CS}^2.
\end{equation}

\begin{figure}
    \centering
    \includegraphics[width=0.4\textwidth]{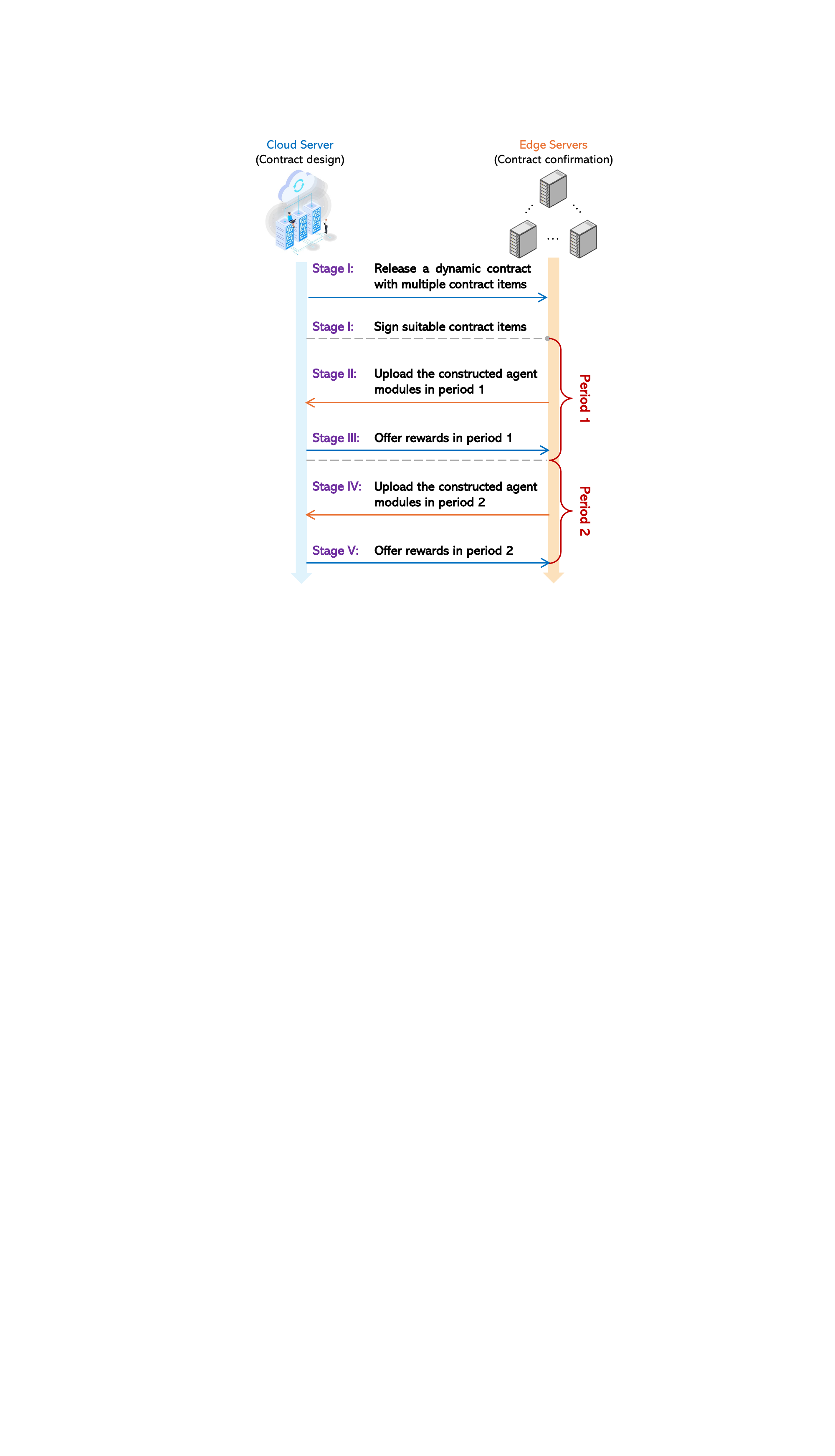}
    \caption{The workflow of the proposed two-period dynamic contract for agent module construction. To minimize disruptions to the construction progress, the proposed dynamic contract remains valid across both periods without requiring redefinition and retransmission to the ESs in period 2.}
    \label{Dynamic_contract}
\end{figure}

The dynamic contract $\Omega$ is directly applied across two periods. According to~\cite{DCTITS}, the long-term contract cycle is divided into five stages, as shown in Fig. \ref{Dynamic_contract}.
\begin{itemize}
    \item \textbf{\textit{Stage I. Contract confirmation:}} At the beginning of federated AI agent construction, the cloud server releases the two-period dynamic contract $\Omega$ with a set of contract items to ESs. Upon receiving the contract, each ES evaluates its WTP and selects a suitable contract item.
    \item \textbf{\textit{Stage II. Agent module creation of period 1:}} After ESs accept selected contract items, the first period of agent module creation begins. ESs first collect feature data from the surrounding environment through inherent wireless sensors. Based on the collected feature data and the initial values informed by the cloud, ESs create agent modules and upload them to the cloud server for integration.
    \item \textbf{\textit{Stage III. Contract realization of period 1:}} At the end of period 1, the cloud server integrates the created agent modules. If the integration is successful, the cloud server will offer the rewards $R_k^1(\theta_k^1)$ specified in the contract to ESs. Otherwise, the ESs that fail to fulfill the contractual requirements will not obtain the corresponding rewards.
    \item \textbf{\textit{Stage IV. Agent module creation of period 2:}} After completing the agent module creation of period 1, ESs can learn their second-period types, and the subsequent process of agent module creation is similar to Stage II.
    \item \textbf{\textit{Stage V. Contract realization of period 2:}} The contract realization of period 2 is similar to Stage III.
\end{itemize}

\subsection{Second-Period Contract Design}
Since the second-period contract is independent~\cite{DynamicLim}, we first use the standard static contract approach~\cite{JinboDiffIoT} to solve the second-period contract. For a feasible contract, the second-period contract should simultaneously satisfy both the IR and IC constraints~\cite{DynamicLim, DCTITS}.
\begin{definition}\label{definition1}
    (IR Constraints in Period 2): Each type-$k$ ES in period 2 can obtain a non-negative utility by selecting the contract item suitable for its second-period type $i$, i.e.,
    \begin{equation}\label{IR2}
        u_i^2(T_i^2(T_k^1)) = \theta_i^2 R_i^2(\theta_k^1) - c T_i^2(\theta_k^1) - E \geq 0,\: \forall i \in \mathcal{K}.
    \end{equation}
\end{definition}
\begin{definition}
    (IC Constraints in Period 2): Each type-$k$ ES in period 2 can achieve the maximum utility by selecting the contract item designed for its second-period type $i$ rather than any other contract item, i.e.,
    \begin{equation}\label{IC2}
        \theta_i^2 R_i^2 - c T_i^2 - E \geq \theta_i^2 R_j^2 - c T_j^2 - E,\: \forall i \neq j \in \mathcal{K}.
    \end{equation}
\end{definition}

IR constraints in (\ref{IR2}) motivate ESs to participate in federated AI agent construction by guaranteeing that their utilities are non-negative if they select suitable contract items. IC constraints in (\ref{IC2}) encourage ESs to reveal their actual type, avoiding the risk that ESs misrepresent their actual types to cheat for higher rewards. In the following, we derive the necessary conditions of the second-period contract.

\begin{lemma}\label{lemma1}
    For any feasible contract, $R_i^2(\theta_k^1) \geq R_j^2(\theta_k^1)$ if and only if $T_i^2(\theta_k^1) \geq T_j^2(\theta_k^1)$, where $i\neq j$.
    \begin{proof}
    Please refer to Appendix A.
\end{proof}
\end{lemma}

\begin{lemma}\label{lemma2}
    For any feasible contract, if $\theta_i^2(\theta_k^1) \geq \theta_j^2(\theta_k^1)$, then $R_i^2(\theta_k^1) \geq R_j^2(\theta_k^1)$.
    \begin{proof}
    Please refer to Appendix B.
    \end{proof}
\end{lemma}

Lemma \ref{lemma1} ensures that ESs can receive higher rewards with more training rounds for improving the model quality of agent modules. Lemma \ref{lemma2} indicates that ESs with a higher WTP level can obtain higher rewards. According to Lemmas \ref{lemma1} and \ref{lemma2}, we can prove the monotonicity property of the second-period contract as follows:
\begin{lemma}\label{lemma3}
    (Monotonicity Property): For any feasible contract, if $\theta_i^2(\theta_k^1) \geq \theta_j^2(\theta_k^1)$, then $T_i^2(\theta_k^1) \geq T_j^2(\theta_k^1)$.
\end{lemma}

Since it is exceedingly challenging to make the dynamic contract solution under $K$ number of IR constraints and $K(K-1)$ number of IC constraints, we reduce the number of IR and IC constraints to derive a tractable set of constraints to solve the second-period contract.
\begin{lemma}\label{lemma4}
    (Reduce IR Constraints in Period 2): The $K$ number of IR constraints in period 2 can be reduced to a single constraint, i.e.,
    \begin{equation}\label{reduced_IR2}
        \theta_1^2 R_1^2(\theta_k^1) - c T_1^2(\theta_k^1) - E = 0.
    \end{equation}
    \begin{proof}
       Please refer to Appendix C.
    \end{proof}
\end{lemma}

\begin{lemma}\label{lemma5}
    The IC constraints in period 2 can be reduced into Local Downward Incentive Compatibility (LDIC) (\ref{LDIC}) and Local Upward Incentive Compatibility (LUIC) (\ref{LUIC}), i.e.,
    \begin{equation}\label{LDIC}
        \theta_i^2 R_i^2(\theta_k^1) - c T_i^2(\theta_k^1) - E \geq \theta_i^2 R_{i-1}^2(\theta_k^1) - c T_{i-1}^2(\theta_k^1) - E.
    \end{equation}
    \begin{equation}\label{LUIC}
        \theta_i^2 R_i^2(\theta_k^1) - c T_i^2(\theta_k^1) - E \geq \theta_i^2 R_{i+1}^2(\theta_k^1) - c T_{i+1}^2(\theta_k^1) - E.
    \end{equation}
    \begin{proof}
        Please refer to Appendix D.
    \end{proof}
\end{lemma}

\begin{lemma}\label{lemma6}
    (Reduce IC Constraints in Period 2): The $K(K-1)$ number of IC constraints in period 2 can be reduced to $(K-1)$ number of IC constraints, i.e.,
    \begin{equation}\label{reduced_IC2}
     \theta_{i}^2 R_i^2 - c T_i^2 - E =  \theta_{i}^2 R_{i-1}^2 - c T_{i-1}^2 - E,\:\forall i \geq 2.
    \end{equation}
    \begin{proof}
    Please refer to Appendix E.
    \end{proof}
\end{lemma}

Based on Lemmas \ref{lemma4} and \ref{lemma6}, the $K$ number of IR constraints (\ref{IR2}) and $K(K-1)$ number of IC constraints (\ref{IC2}) can be reduced to one IR constraint (\ref{reduced_IR2}) and $(K-1)$ number of IC constraints (\ref{reduced_IC2}). We derive the sufficient and necessary conditions of the feasible contract in period 2 as follows:
\begin{theorem}\label{theorem1}
    (Sufficient and Necessary Conditions in Period 2): A feasible contract in period 2 must satisfy the following sufficient and necessary conditions:
    \begin{align*}
     &1)\quad 0 \leq R_{1}^2(\theta_k^1) \leq \cdots \leq R_{k}^2(\theta_k^1) \leq \cdots \leq R_{K}^2(\theta_k^1). \\
     &2)\quad 0 \leq T_{1}^2(\theta_k^1) \leq \cdots \leq T_{k}^2(\theta_k^1) \leq \cdots \leq T_{K}^2(\theta_k^1). \\
     &3)\quad \theta_1^2 R_1^2(\theta_k^1) - c T_1^2(\theta_k^1) - E = 0. \\
     &4)\quad \theta_{i}^2 R_i^2 - c T_i^2 - E =  \theta_{i}^2 R_{i-1}^2 - c T_{i-1}^2 - E,\:\forall i \geq 2.
    \end{align*}
\end{theorem}
In Theorem \ref{theorem1}, the first two conditions indicate the monotonicity property of the second-period contract. The last two conditions are the simplified IR and IC constraints in period 2, respectively. Based on (\ref{reduced_IR2}) and (\ref{reduced_IC2}), the optimal reward in period 2 can be derived by using the iterative method.
\begin{theorem}\label{theorem2}
    (Optimal Rewards in Period 2): The optimal reward $(R_i^2)^{\star}$ in period 2 is given by
    \begin{equation}\label{optimalreward2}
        (R_i^2)^{\star} = \left\{
        \begin{aligned}
            & \frac{cT_{1}^2(\theta_k^1) + E}{\theta_1^2},\: i = 1,\\
            & \frac{cT_{1}^2(\theta_k^1) + E}{\theta_1^2} + c \sum_{j=2}^i \frac{T_{j}^2(\theta_k^1) - T_{j-1}^2(\theta_k^1)}{\theta_j^2},\:i \neq 1.
        \end{aligned}
        \right.  
    \end{equation}
    \begin{proof}
    Please refer to Appendix F.
    \end{proof}
\end{theorem}

\subsection{First-Period Contract Design}
For the feasible contract design in period 1, the IR constraints are similar to those in Definition \ref{definition1}, as given by
\begin{equation}
u_k^1(T_k^1) = \theta_k^1 R_k^1 - c T_k^1 - E \geq 0,\: \forall k \in \mathcal{K}.
\end{equation}

Since ESs may misreport their WTP levels in period 1 to gain higher utilities in period 2~\cite{DCTITS}, instead of considering just the one-period IC constraints, the cloud server should consider the Intertemporal IC (IIC) constraints to incentivize ESs to reflect their actual types~\cite{DCTITS}, and the IIC constraints are expressed as
\begin{equation}\label{IIC}
\begin{split}
   &u_k^1(T_k^1) +  \beta \sum_{j=1}^K p_j^2(T_k^1) u_j^2(T_j^2(T_k^1))\geq\\
   & u_k^1(T^1_{k^{\prime}}) + \beta \sum_{j=1}^K p_j^2(T_k^1) u_j^2(T_j^2(T^1_{k^{\prime}})),\: \forall k \neq k^{\prime} \in \mathcal{K}.
\end{split}
\end{equation}

From the IIC constraints in (\ref{IIC}), we know that ESs will obtain higher utilities by truthfully declaring their WTP levels. When the first-period and second-period types are mutually independent, we have $\sum_{j=1}^K p_j^2(T_k^1) u_j^2(T_j^2(T_k^1)) = \sum_{j=1}^K p_j^2(T_k^1) u_j^2(T_j^2(T^1_{k^{\prime}}))$, where $\sum_{j=1}^K p_j^2(T_k^1) u_j^2(T_j^2(T^1_{k^{\prime}}))$ represents the discounted expected utility of the type-$k$ ESs that declare their types are $\theta^1_{k^{\prime}}$ in period 1, and the IIC constraints can be simplified as the IC constraints of the standard static contract. In the following, we derive the necessary conditions for the feasible contract in period 1.
\begin{lemma}\label{lemma7}
    For any feasible contract in period 1, if $\theta_k^1 > \theta^1_{k^{\prime}}$, then $T_k^1 > T^1_{k^{\prime}}$, and $T_j^2(T_k^1) > T_j^2(T^1_{k^{\prime}})$, where $k\neq k^{\prime}$.
    \begin{proof}
    Please refer to Appendix G.
    \end{proof}
\end{lemma}

\begin{lemma}\label{lemma8}
    For any feasible contract in period 1, $T_k^1 > T^1_{k^{\prime}}$ if and only if $R_k^1 > R^1_{k^{\prime}}$, where $k \neq k^{\prime}$.
    \begin{proof}
    Please refer to Appendix H.
    \end{proof}
\end{lemma}

Lemma \ref{lemma7} indicates that the ESs with higher WTP levels will perform more training rounds for agent module creation in both period 1 and period 2, even if they have the same WTP level in period 2. Lemma \ref{lemma8} represents the monotonicity of first-period rewards and training rounds, indicating that the ESs with higher WTP levels will receive higher rewards from the cloud server in period 1. Next, we reduce the IR and IIC constraints in period 1, which is similar to period 2.

\begin{lemma}\label{lemma9}
    (Reduce IR Constraints in Period 1): Similar to (\ref{reduced_IR2}), the $K$ number of IR constraints in period 1 can also be reduced to a single constraint, i.e.,
    \begin{equation}\label{reduced_IR1}
    \theta_1^1 R_1^1 - cT_1^1 - E = 0.
    \end{equation}
\end{lemma}

\begin{lemma}\label{lemma10}
    (Reduce IIC Constraints in Period 1): Similar to (\ref{reduced_IC2}), the $K(K-1)$ number of IIC constraints can be reduced into the $(K-1)$ number of local downward IIC constraints, which is given by
    \begin{equation}\label{reduced_IIC}
    \begin{split}
    u_k^1(T_k^1) &+ \beta\sum_{j=1}^K p_j^2(T_k^1)u_j^2(T_j^2(T_k^1))\\
    &= u_k^1(T^1_{k-1}) + \beta\sum_{j=1}^K p_j^2(T_k^1) u_j^2(T_j^2(T^1_{k-1})). 
    \end{split}
    \end{equation}
\end{lemma}

The proofs of Lemmas \ref{lemma9} and \ref{lemma10} are similar to Lemmas \ref{lemma4} and \ref{lemma6}, respectively. We present the sufficient and necessary conditions of the feasible contract in period 1 in the following.

\begin{theorem}\label{theorem3}
  (Sufficient and Necessary Conditions in Period 1): A feasible contract in period 1 must satisfy the following sufficient and necessary conditions:
        \begin{align*}
            &1)\quad 0 \leq R_1^1 \leq \cdots \leq R_{k}^1 \leq \cdots \leq R_{K}^1,\\
            &2)\quad 0 \leq T_{1}^1 \leq \cdots \leq T_{k}^1 \leq \cdots \leq T_{K}^1,\\
            &3)\quad \theta_1^1 R_1^1 - cT_1^1 - E = 0,\\
            &4)\quad u_k^1(T_k^1) + \beta\sum_{j=1}^K\nolimits p_j^2(T_k^1)u_j^2(T_j^2(T_k^1))\notag \\
            &\qquad\quad\:= u_k^1(T^1_{k-1}) + \beta\sum_{j=1}^K\nolimits p_j^2(T_k^1) u_j^2(T_j^2(T^1_{k-1})).
        \end{align*}
\end{theorem}

In Theorem \ref{theorem3}, the first two conditions are the monotonicity property of the first-period contract. The last two conditions present the simplified IR and IIC constraints in period 1, respectively. Based on (\ref{reduced_IR1}) and (\ref{reduced_IIC}), the optimal reward in period 1 can also be derived using the iterative method.
\begin{theorem}\label{theorem4}
    (Optimal Rewards in Period 1): The optimal reward $(R_i^1)^{\star}$ in period 1 is given by
    \begin{equation}\label{optimalreward1}
        (R_i^1)^{\star} = \left\{
        \begin{aligned}
            & \frac{cT_{1}^1 + E}{\theta_1^1},\: i = 1,\\
            & \frac{cT_{1}^1 + E}{\theta_1^1} + \sum_{k=2}^i\Bigg[(T_k^1-T^1_{k-1}) \frac{c}{\theta_k^1}\\
            &\qquad\qquad\quad+ \frac{\beta}{\theta_k^1}\sum_{j=2}^K p_j^2(T_k^1)\Phi_j\Bigg],\:i \neq 1,
        \end{aligned}
        \right.  
    \end{equation}
where $\Phi_j = \sum_{l=1}^{j-1}\Big[\theta_j^2c(T_l^2(\theta^1_{k-1})-T_l^2(\theta_k^1))\big(\frac{1}{\theta_l^2}-\frac{1}{\theta^2_{l+1}}\big)\Big]$.
\begin{proof}
    The proof of the optimal reward (\ref{optimalreward1}) in period 1 is similar to the proof of the optimal reward (\ref{optimalreward2}) in period 2. 
\end{proof}
\end{theorem}

\subsection{Optimal Two-Period Contract}
To solve the optimal two-period contract, we substitute the derived optimal rewards (\ref{optimalreward1}) and (\ref{optimalreward2}) into the expected utility of the cloud server (\ref{expected_utility}), and the expected utility of the cloud server in period 2 is expressed as
\begin{equation}
    U_{CS}^2(T_j^2(\theta_k^1)) = \Bigg[\sum_{j=1}^K p_j^2(T_k^1)N(\alpha Q_j^2 - R_j^2)\Bigg]- \sigma E_C,
\end{equation}
where $Q_j^2 = 1 - 2^\frac{-T_j^2(\theta_k^1)(2-L\delta)\delta \varepsilon}{2}$.

Similarly, the expected utility of the cloud server in period 1 is expressed as
\begin{equation}
    U_{CS}^1(T_k^1) = \Bigg[\sum_{k=1}^K p_k^1(T_k^1)N(\alpha Q_k^1 - R_k^1)\Bigg]- \sigma E_C,
\end{equation}
where $Q_k^1 = 1 - 2^\frac{-T_k^1(2-L\delta)\delta \varepsilon}{2}$.

Therefore, based on (\ref{total_profit}), the dynamic contract optimization problem is formulated as
\begin{equation}\label{problem1}
    \begin{split}
        & \max_{\{T_k^1, T_j^2(\theta_k^1)\}} U_{CS}^1(T_k^1) + \beta U_{CS}^2(T_j^2(\theta_k^1)) \\
        &\quad\:\:\:\rm{s.t.}\quad \text{(\ref{optimalreward2})}\:\: \text{and}\:\: \text{(\ref{optimalreward1})}.
    \end{split}
\end{equation}

\begin{figure*}[t]
    \centering
    \vspace{-0.5cm}
    \includegraphics[width=0.9\textwidth]{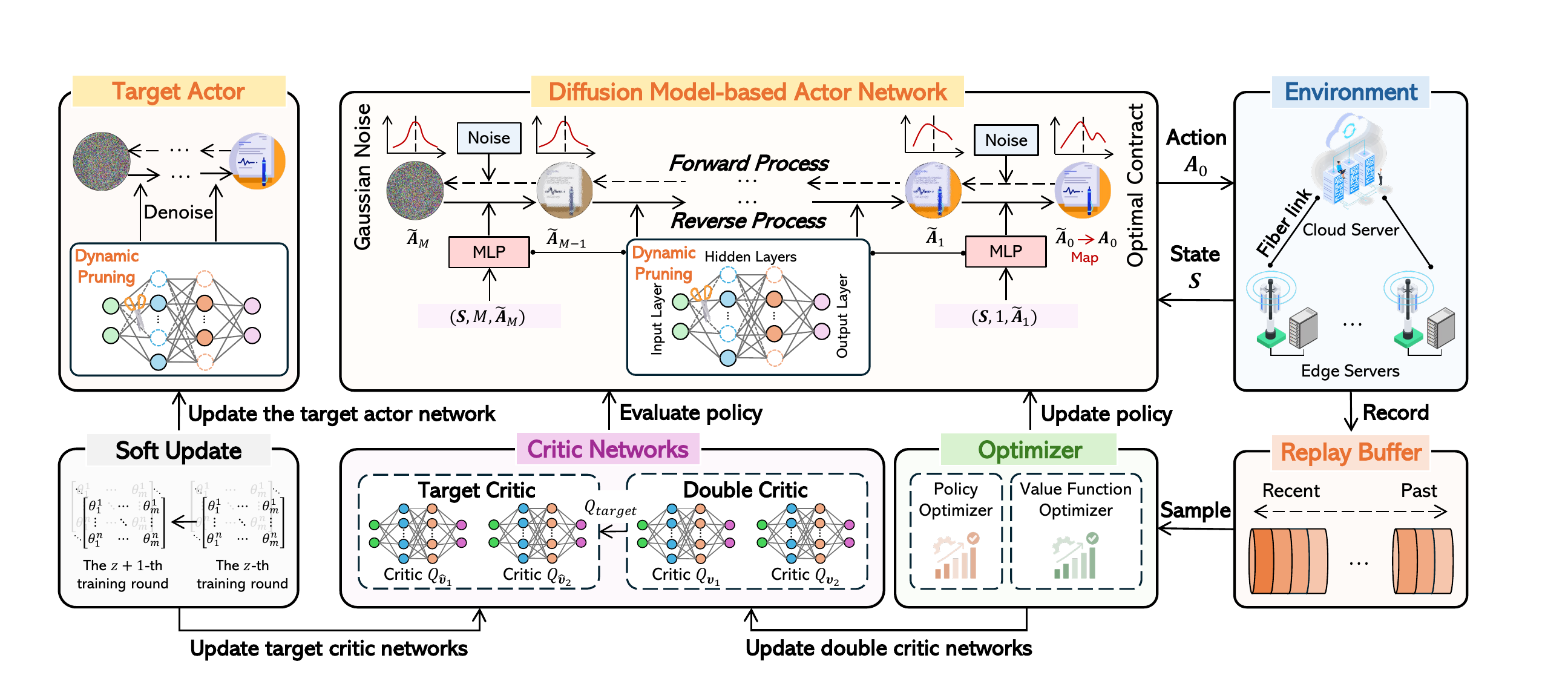}
    \caption{The architecture of the proposed EDMSAC algorithm for dynamic contract design. The EDMSAC algorithm can generate optimal two-period contracts through the denoising process. To enhance denoising efficiency, we innovatively apply dynamic structured pruning techniques to the MLP of DM-based actor networks. Note that the unimportant neurons in the actor networks are simply masked, not removed outright.}
    \label{EDMSAC_algorithm}
\end{figure*}

\begin{remark}
The number of variables in the dynamic contract optimization problem (\ref{problem1}) has been reduced from $4K$ to $2K$. Currently, researchers rely on traditional optimization algorithms, such as heuristic algorithms, to solve this problem~\cite{DCTITS, DynamicLim}. However, in practical scenarios, the parameters of ESs and the cloud server, such as communication channel conditions, fluctuate dynamically. As a result, traditional optimization algorithms must be frequently redesigned and re-implemented, leading to substantial computing resource consumption. Furthermore, traditional algorithms often fail to generalize to practical scenarios when the environment lacks complete and accurate information~\cite{wen2024sustainablediffusionbasedincentivemechanism}. DRL algorithms have been successfully applied for optimal static contract design due to their adaptability to dynamic environments~\cite{10829636, 10368063}. In particular, PPO, as an on-policy and model-free algorithm, iteratively updates the policy towards optimal performance of contract generation by using a clipped surrogate objective~\cite{wen2024learningbasedbigdatasharing}, and SAC, as an off-policy algorithm, trains a stochastic policy for optimal dynamic contract generation by maximizing both the expected cumulative reward and policy entropy~\cite{JinboDiffIoT}. Notably, researchers have adopted GDMs to identify optimal static contracts, achieving superior performance compared with traditional DRL algorithms~\cite{JinboDiffIoT, wen2024sustainablediffusionbasedincentivemechanism}. In particular, the DMSAC algorithm can improve stability and performance over baseline DRL algorithms~\cite{10409284}. Building on this foundation~\cite{10409284}, we propose the EDMSAC algorithm that leverages lightweight GDMs to generate optimal dynamic contracts, effectively adapting to the changing environment during federated AI agent construction.
\end{remark}

\section{Enhanced Diffusion Model-based Soft Actor-Critic Algorithms for Optimal Dynamic Contract Design}\label{EDMSAC}
In this section, we first model the optimization problem (\ref{problem1}) as a Markov Decision Process (MDP). Subsequently, we introduce the Denoising Diffusion Probabilistic Model (DDPM). Finally, we present the enhanced diffusion policy for optimal dynamic contract design, as shown in Fig. \ref{EDMSAC_algorithm}.

\subsection{Markov Decision Process Formulation}
MDP is a mathematical framework for modeling decision-making problems~\cite{moghaddasi2024enhanced}, which is represented as a five-element tuple consisting of a state space, an action space, a reward function, a discount factor determining the weight of the future reward relative to the current reward, and state transition probabilities. In the following, we explicitly define the state space, action space, and reward function of the MDP framework.
\subsubsection{State space} At each step $z \in \{1,2,\ldots, Z\}$, the state space from the environment is defined as
\begin{equation}
    \boldsymbol{S} \triangleq\{N, K, \sigma, P_n, r_{n,c}, E_C, p_k^1, p_j^2(T_k^1), \theta_k^1, \theta_j^2\},\:\forall j,k\in \mathcal{K},
\end{equation}
where $N$ and $K$ are constants that do not evolve over time, while the other terms vary with each training round. Notably, the size of the state space $\boldsymbol{S}$ is $(6 + 3K + K^2)$.

\subsubsection{Action space} According to the optimization problem (\ref{problem1}), the action space in the $z$-th training round is defined as
\begin{equation}
    \boldsymbol{A} \triangleq \{T_k^1, T_j^2(\theta_k^1)\},\:\forall j,k\in \mathcal{K},
\end{equation}
where $\boldsymbol{A}$ is mapped from the denoising action $\Tilde{\boldsymbol{A}}$.

\subsubsection{Reward function} After executing the action $\boldsymbol{A}$ based on the current state $\boldsymbol{S}$, the DRL agent receives an immediate reward $r(\boldsymbol{S}, \boldsymbol{A})$. To enhance the stability of its learning process, we define a penalty-based reward function as
\begin{equation}\label{DRL_Reward}
r(\boldsymbol{S}, \boldsymbol{A}) = U_{CS}^1(T_k^1) + \beta U_{CS}^2(T_j^2(\theta_k^1)) - \underbrace{\lambda \sum_{i=1}^{2K} (\Tilde{\boldsymbol{A}}[i])^2}_{\rm{Action}\:\rm{penalty}},
\end{equation}
where $\lambda$ is a pre-defined parameter that controls the degree of the action penalty.

\subsection{Denoising Diffusion Probabilistic Models}
DDPMs are powerful generative models that generate data based on forward and reverse processes~\cite{wang2023diffusionpoliciesexpressivepolicy}. In the forward process, given initial data $\boldsymbol{x}_0 \sim q(\boldsymbol{x}_0)$, the Gaussian noise is gradually added to the data $\boldsymbol{x}_0 \sim q(\boldsymbol{x}_0)$ across $M$ steps with a pre-defined variance schedule $\epsilon$~\cite{NEURIPS2020_4c5bcfec}, described as
\begin{equation}
    q(\boldsymbol{x}_m|\boldsymbol{x}_{m-1}) := \mathcal{N}(\boldsymbol{x}_m; \sqrt{1-\epsilon_m}\boldsymbol{x}_{m-1}, \epsilon_m\mathbf{I}),
\end{equation}
where $\epsilon_m = 1 - e^{-\frac{\epsilon_{min}}{M}-\frac{2m-1}{2M^2}(\epsilon_{max} - \epsilon_{min})}$ and $\mathbf{I}$ is the identity matrix. By using the Markov chain property, the joint distribution of $\boldsymbol{x}_{1},\ldots,\boldsymbol{x}_{M}$ conditioned on $\boldsymbol{x}_{0}$ is given by
\begin{equation}
    q(\boldsymbol{x}_{1:M}|\boldsymbol{x}_{0}) = \prod_{m=1}^M q(\boldsymbol{x}_m|\boldsymbol{x}_{m-1}).
\end{equation}

In the reverse process, the DDPM denoises the noisy data $\boldsymbol{x}_{M}$ by learning parameterized transitions $g_{\boldsymbol{\theta}}(\boldsymbol{x}_{m-1}|\boldsymbol{x}_{m})$~\cite{NEURIPS2020_4c5bcfec}, which is expressed as
\begin{equation}\label{prediction_model}
    g_{\boldsymbol{\theta}}(\boldsymbol{x}_{m-1}|\boldsymbol{x}_{m}) = \mathcal{N}(\boldsymbol{x}_{m-1};\mu_{\boldsymbol{\theta}}(\boldsymbol{x}_m,m),\Sigma_{\boldsymbol{\theta}}(\boldsymbol{x}_m,m)),
\end{equation}
where $\mu_{\boldsymbol{\theta}}$ and $\Sigma_{\boldsymbol{\theta}}$ are parameterized by deep neural networks with model parameters $\boldsymbol{\theta}$. Hence, the trajectory from $\boldsymbol{x}_M$ to $\boldsymbol{x}_0$ under the condition that $\prod_{m=1}^M(1-\epsilon_m)\approx 0$ is given by
\begin{equation}\label{reverse_process}
    g_{\boldsymbol{\theta}}(\boldsymbol{x}_{0:M}) = \mathcal{N}(\boldsymbol{x}_M;\mathbf{0},\mathbf{I})\prod_{m=1}^M g_{\boldsymbol{\theta}}(\boldsymbol{x}_{m-1}|\boldsymbol{x}_{m}).
\end{equation}

The training objective of DDPMs is to maximize the variational lower bound defined as $\mathbb{E}_{q(\boldsymbol{x}_{0:M})}\big[\ln \frac{g_{\boldsymbol{\theta}}(\boldsymbol{x}_{0:M})}{q(\boldsymbol{x}_{1:M}|\boldsymbol{x}_{0})}\big]$~\cite{wang2023diffusionpoliciesexpressivepolicy}. After completing the training of DDPMs, the original data $\boldsymbol{x}_{0}$ can be recovered from the noisy data $\boldsymbol{x}_{M}$.

\begin{algorithm}[t]
\label{diffusion_algorithm}
\DontPrintSemicolon
\SetAlgoLined

Initialize policy network $\pi_{\boldsymbol{\theta}}$, critic networks $Q_{\boldsymbol{\upsilon}_1}, Q_{\boldsymbol{\upsilon}_2}$, target networks $\hat{\pi}_{\hat{\boldsymbol{\theta}}}, Q_{\hat{\boldsymbol{\upsilon}}_1}, Q_{\hat{\boldsymbol{\upsilon}}_2}$, relay buffer $\mathcal{D}$, pruning rate $\varrho$, and mask matrices $\boldsymbol{W}$.

\textcolor{blue}{\#\# \textbf{\textit{Training}}}

\For{\rm{the training step} $v=1$ \rm{to} $V$}
{
\For{\rm{the number of transitions} $z=1$ \rm{to} $Z$}
{
\textcolor{blue}{\#\#\# \textit{Generating contracts}}

    Observe state $\boldsymbol{S}_z$ and initialize a random normal distribution $\Tilde{\boldsymbol{A}}_M \sim \mathcal{N}(\mathbf{0}, \mathbf{I})$.

    Map $\Tilde{\boldsymbol{A}}_0$ to $\boldsymbol{A}_0$ and perform $\boldsymbol{A}_0$, observe the next state $\boldsymbol{S}_{z+1}$, and receive reward $r_z$. 

    Store record $(\boldsymbol{S}_z,\boldsymbol{A}_0,\boldsymbol{S}_{z+1}, r_z)$ into $\mathcal{D}$.

}
    
    \textcolor{blue}{\#\#\# \textit{Dynamic pruning}}
    
    Calculate neuron importance of $\boldsymbol{\theta}$ and $\hat{\boldsymbol{\theta}}$ by using (\ref{importance}), respectively.

    Use the mask matrices $\boldsymbol{W}$ to mask output neurons based on the pruning rate $\varrho$ by (\ref{remove_neuron}).

    \textcolor{blue}{\#\#\# \textit{Parameter updates}}

    Sample a random mini-batch of transitions $\mathcal{B}_z$ with size $B$ from $\mathcal{D}$.

    Update the policy $\pi_{\boldsymbol{\theta}}$ using $\mathcal{B}_z$ by (\ref{policy_update}).
    
    Update $Q_{\boldsymbol{\upsilon}_1}, Q_{\boldsymbol{\upsilon}_2}$ using $\mathcal{B}_z$ by (\ref{Q_update}).

    Update target networks $\hat{\pi}_{\hat{\boldsymbol{\theta}}}, Q_{\hat{\boldsymbol{\upsilon}}_1}, Q_{\hat{\boldsymbol{\upsilon}}_2}$ by (\ref{targets}).

}
Reconstruct the compact policy networks.

\textcolor{blue}{\#\# \textbf{\textit{Inference}}}

    Input a state $\boldsymbol{S}$.
    
    Generate $\Tilde{\boldsymbol{A}}_0$ based on the target policy $\hat{\pi}_{\hat{\boldsymbol{\theta}}}$ by (\ref{generate_contract}).
    
    \textbf{return} $\boldsymbol{A}_0=\Omega^{\star}$.
\caption{EDMSAC for Dynamic Contract Design}\label{diffusion_algorithm}
\end{algorithm}

\subsection{Enhanced Diffusion Policy for Dynamic Contract Design}

We define the \textit{dynamic contract design policy} as $\pi_{\boldsymbol{\theta}}(\Tilde{\boldsymbol{A}}|\boldsymbol{S})$ with parameters $\boldsymbol{\theta}$. The goal of $\pi_{\boldsymbol{\theta}}(\Tilde{\boldsymbol{A}}|\boldsymbol{S})$ is to generate optimal dynamic contracts by maximizing the expected cumulative reward based on (\ref{DRL_Reward}), which is given by~\cite{10409284}
\begin{equation}
    \mathbb{E} \Bigg[\sum_{z = 0} ^ Z \gamma^z (r(\boldsymbol{S}_z, \boldsymbol{A}_z)+\kappa \mathcal{H}(\pi_{\boldsymbol{\theta}}(\cdot|\boldsymbol{S}_z)))\Bigg],
\end{equation}
where $\gamma$ is the discount factor for future rewards, $\mathcal{H}(\pi_{\boldsymbol{\theta}}(\cdot|\boldsymbol{S}_z))$ represents the entropy of the policy $\pi_{\boldsymbol{\theta}}(\boldsymbol{S}_z)$, and $\kappa$ is the temperature coefficient that controls the strength of the entropy.

According to (\ref{reverse_process}), the dynamic contract design policy $\pi_{\boldsymbol{\theta}}(\Tilde{\boldsymbol{A}}|\boldsymbol{S})$ can be obtained through the reverse process of a conditional DM, expressed as~\cite{ wang2023diffusionpoliciesexpressivepolicy}
\begin{equation}
   \pi_{\boldsymbol{\theta}}(\Tilde{\boldsymbol{A}}|\boldsymbol{S}) = \mathcal{N}(\Tilde{\boldsymbol{A}}_M;\mathbf{0},\mathbf{I})\prod_{m=1}^M g_{\boldsymbol{\theta}}(\Tilde{\boldsymbol{A}}_{m-1}|\Tilde{\boldsymbol{A}}_{m}, \boldsymbol{S}),
\end{equation}
where the end sample of the reverse chain is $\Tilde{\boldsymbol{A}}_0$. According to~\cite{wang2023diffusionpoliciesexpressivepolicy}, $g_{\boldsymbol{\theta}}(\Tilde{\boldsymbol{A}}_{m-1}|\Tilde{\boldsymbol{A}}_{m}, \boldsymbol{S})$ is modeled as a Gaussian distribution $\mathcal{N}(\Tilde{\boldsymbol{A}}_{m-1};\mu_{\boldsymbol{\theta}}(\Tilde{\boldsymbol{A}}_m, \boldsymbol{S}, m),\Sigma_{\boldsymbol{\theta}}(\Tilde{\boldsymbol{A}}_m,\boldsymbol{S},m))$ based on (\ref{prediction_model}), with the covariance matrix $\Sigma_{\boldsymbol{\theta}}(\Tilde{\boldsymbol{A}}_m,\boldsymbol{S},m) = \frac{\zeta_m(1 - \bar{\chi}_{m-1})}{1 - \bar{\chi}_m} \mathbf{I}$ and the mean $\mu_{\boldsymbol{\theta}}(\Tilde{\boldsymbol{A}}_m, \boldsymbol{S}, m)$ constructed as
\begin{equation}
    \mu_{\boldsymbol{\theta}}(\Tilde{\boldsymbol{A}}_m, \boldsymbol{S}, m)=\frac{1}{\sqrt{\chi_m}}\bigg(\Tilde{\boldsymbol{A}}_m-\frac{\zeta_m\tanh{\xi_{\boldsymbol{\theta}}(\Tilde{\boldsymbol{A}}_m,\boldsymbol{S},m)}}{\sqrt{1-\bar{\chi}_m}}\bigg),
\end{equation}
where $\zeta_m \in (0,1)$ is a hyperparameter for model training, $\chi_m = 1 - \zeta_m$, and $\bar{\chi}_m = \prod_{i=0}^m \chi_i$~\cite{wen2024sustainablediffusionbasedincentivemechanism, wang2023diffusionpoliciesexpressivepolicy}. $\xi_{\boldsymbol{\theta}}(\Tilde{\boldsymbol{A}}_m,\boldsymbol{S},m)$ is the Multi-Layer Perceptron (MLP) of DM-based actor networks, as shown in Fig. \ref{EDMSAC_algorithm}. We first sample $\Tilde{\boldsymbol{A}}_M \sim \mathcal{N}(\mathbf{0}, \mathbf{I})$, and then a series of samples $\Tilde{\boldsymbol{A}}_{m},\: m = M,\ldots,1$ can be obtained through the reverse diffusion chain parameterized by $\boldsymbol{\theta}$, which is given by
\begin{equation}\label{generate_contract}
 \Tilde{\boldsymbol{A}}_{m-1}|\Tilde{\boldsymbol{A}}_{m} = \frac{\Tilde{\boldsymbol{A}}_{m}}{\sqrt{\chi_m}}-\frac{\zeta_m}{\sqrt{\chi_m(1-\bar{\chi}_m)}}\xi_{\boldsymbol{\theta}}(\Tilde{\boldsymbol{A}}_m,\boldsymbol{S},m)+\sqrt{\zeta_m}\xi,
\end{equation}
where $\xi \sim \mathcal{N}(\mathbf{0},\mathbf{I})$. Notably, when $m = 1$, we have $\xi = 0$~\cite{wang2023diffusionpoliciesexpressivepolicy, NEURIPS2020_4c5bcfec}. For policy improvement, we introduce the Q-function~\cite{wang2023diffusionpoliciesexpressivepolicy}. Specifically, we create two critic networks $Q_{\boldsymbol{\upsilon}_1}, Q_{\boldsymbol{\upsilon}_2}$, target critic networks $Q_{\hat{\boldsymbol{\upsilon}}_1}, Q_{\hat{\boldsymbol{\upsilon}}_2}$, and a target policy $\hat{\pi}_{\hat{\boldsymbol{\theta}}}$. The final policy-learning objective is to minimize the actor loss, which is given by
\begin{equation}
    \mathcal{L}(\boldsymbol{\theta}) = - \mathbb{E}_{\boldsymbol{S}\sim\mathcal{D},\Tilde{\boldsymbol{A}}_{0}\sim \pi_{\boldsymbol{\theta}}} [Q_{\boldsymbol{\upsilon}}(\boldsymbol{S}, \Tilde{\boldsymbol{A}}_{0}) + \kappa \log\pi_{\boldsymbol{\theta}}(\Tilde{\boldsymbol{A}}_{0}|\boldsymbol{S})],
\end{equation}
where $Q_{\boldsymbol{\upsilon}}(\boldsymbol{S}, \Tilde{\boldsymbol{A}}_{0}) = \min\{Q_{\boldsymbol{\upsilon}_1}(\boldsymbol{S}, \Tilde{\boldsymbol{A}}_{0}), Q_{\boldsymbol{\upsilon}_2}(\boldsymbol{S}, \Tilde{\boldsymbol{A}}_{0})\}$ and $\mathcal{D}$ is the relay buffer. To update the Q-function, we minimize the temporal difference error, expressed as
\begin{equation}\label{Q_update}
\begin{split}
    &\mathbb{E}_{(\boldsymbol{S}_z, \boldsymbol{A}_z, \boldsymbol{S}_{z+1}, r_z)\sim \mathcal{B}_z}\Big[\sum_{i=1,2}(r(\boldsymbol{S}_z,\boldsymbol{A}_z)+\gamma^z(1-d_{z+1})\\
    &\:(Q_{\hat{\boldsymbol{\upsilon}}}(\boldsymbol{S}_{z+1}) - \kappa \log \hat{\pi}_{\hat{\boldsymbol{\theta}}}(\boldsymbol{S}_{z+1}))-Q_{\boldsymbol{\upsilon}_i}(\boldsymbol{S}_z,\boldsymbol{A}_z))^2\Big],
\end{split}
\end{equation}
where $\mathcal{B}_z$ is a mini-batch of transitions sampled from $\mathcal{D}$ and $d_{z+1}\in \{0,1\}$ is a terminated flag, with $d_{z+1} = 1$ indicating that the training episode has ended. 

To enhance denoising efficiency, we employ dynamic structured pruning techniques to linear layers of $\xi_{\boldsymbol{\theta}}(\cdot,\cdot,\cdot)$. Considering $\xi_{\boldsymbol{\theta}}(\cdot,\cdot,\cdot)$ with $H$ fully connected layers~\cite{wen2024sustainablediffusionbasedincentivemechanism}, we denote the parameters in the $h$-th fully connected layer as $\boldsymbol{\theta}^{(h)} \in \mathbb{R}^{x \times y},\: h\in\{1,\ldots,H\}$. We first calculate the importance of each row in $\boldsymbol{\theta}^{(h)}$, which is expressed as
\begin{equation}\label{importance}
    ||\boldsymbol{\theta}^{(h)}_i||_2 = \sqrt{\sum_{j=1}^y \theta_{i,j}^2},\: i = 1,2,\ldots,x.
\end{equation}
Then, we introduce mask matrices $\boldsymbol{W}^{(h)}\in \mathbb{R}^{x \times y}$ to prune fully connected layers with low importance based on the pre-defined pruning rate. The elements of $\boldsymbol{W}^{(h)}\in \mathbb{R}^{x \times y}$ are binary, where $\boldsymbol{W}^{(h)}_{x} = 0$ indicates that the output neuron $o_x^{(h)}$ is pruned, while $o_x^{(h)}$ remains otherwise. Hence, the pruned parameters in the $h$-th fully connected layer are given by
\begin{equation}\label{remove_neuron}
\boldsymbol{\theta}'^{(h)} = \boldsymbol{\theta}^{(h)} \odot \boldsymbol{W}^{(h)},
\end{equation}
where $\odot$ represents the Hadamard product~\cite{10416899}. The pruned policy parameters $\boldsymbol{\theta}'$ can be updated by performing gradient descent algorithms such as Adam, as given by
\begin{equation}\label{policy_update}
    \boldsymbol{\theta}'_{v+1} \gets \boldsymbol{\theta}'_{v} - \eta \nabla_{\boldsymbol{\theta}'_{v}}\mathcal{L}(\boldsymbol{\theta}),
\end{equation}
where $\boldsymbol{\theta}'_{v}$ are the pruned policy parameters in the $v$-th training step and $\eta \in (0,1]$ is the learning rate of the actor. Similarly, the parameters of the target policy are also dynamically pruned. During gradient descent, the parameters of the target networks remain frozen and then are updated through a soft update mechanism~\cite{wang2023diffusionpoliciesexpressivepolicy, 10409284}, which is given by
\begin{equation}\label{targets}
\begin{split}
    \hat{\boldsymbol{\theta}}'_{v+1} &\gets \tau \boldsymbol{\theta}'_v + (1-\tau)\hat{\boldsymbol{\theta}}'_{v},\\
    \hat{\boldsymbol{\upsilon}}_{i, v+1} &\gets \tau \boldsymbol{\upsilon}_{i, v} + (1-\tau)\hat{\boldsymbol{\upsilon}}_{i, v},\: \text{for}\: i = \{1,2\},
\end{split}  
\end{equation}
where $\boldsymbol{\upsilon}_{i, v}$ gives the critic parameters in the $v$-th training step and $\tau \in (0,1]$ is the update rate of the target networks.
\begin{figure*}[t]
    \centering
    \vspace{-0.5cm}
    \includegraphics[width=0.9\textwidth]{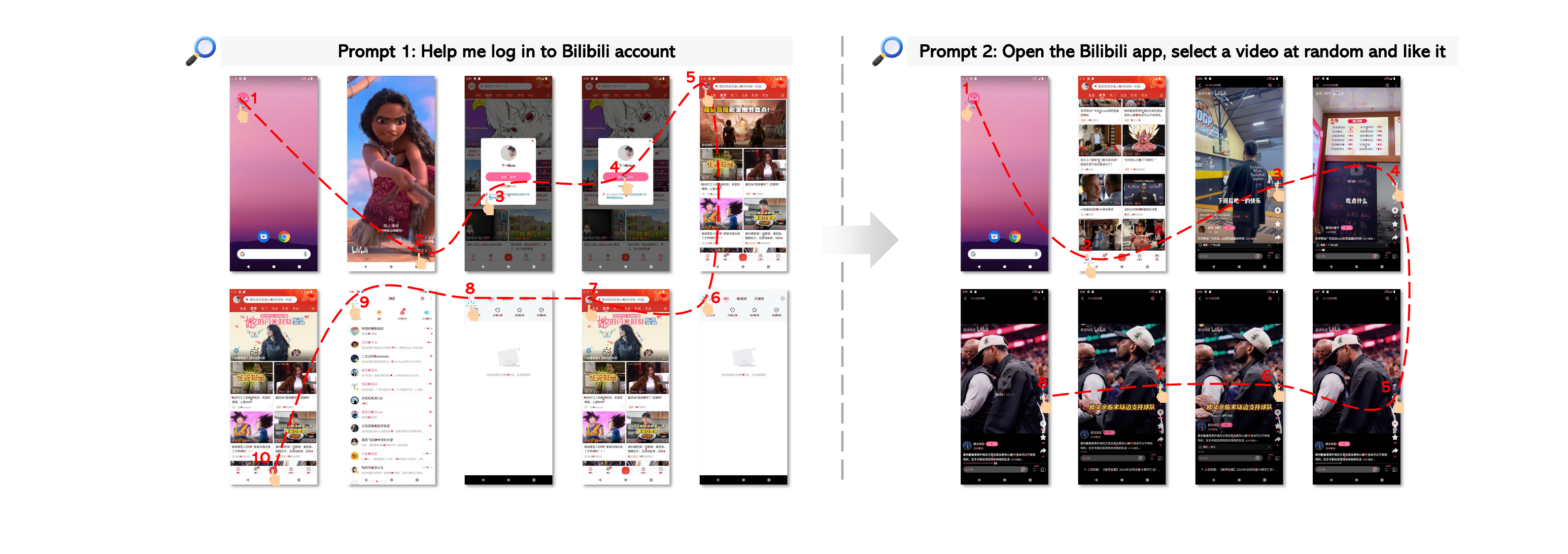}
    \caption{Two case studies of mobile phone operations based on an AI agent. We call the Qwen-VL-Plus and Qwen-VL-Max models through the Qwen-VL API to construct the mobile AI agent, utilizing Android Debug Bridge (ADB) to operate a mobile phone with the Android OS system, where ADB can simulate all operations of the AI agent~\cite{wang2024mobileagentv2mobiledeviceoperation}.}
    \label{case_study}
\end{figure*}

\begin{figure}[t]
	\centering
	\subfigure[ Case 1.]
	{\includegraphics[width=0.24\textwidth]{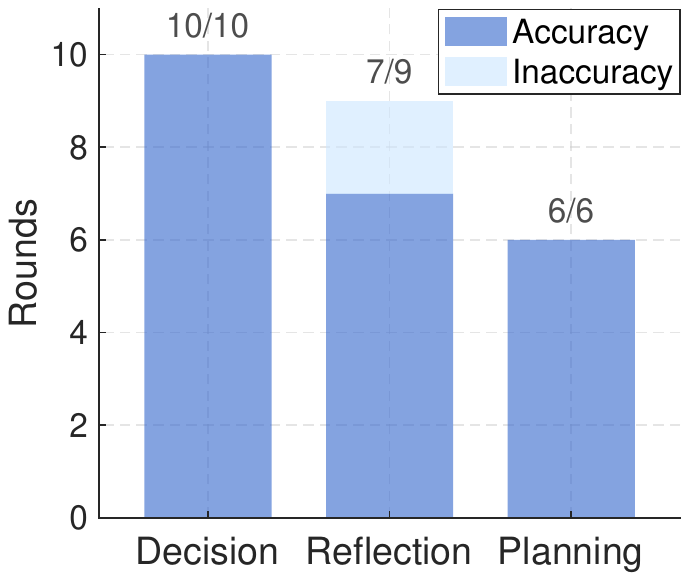}\label{Case_1}}
	\subfigure[ Case 2.]
	{\includegraphics[width=0.235\textwidth]{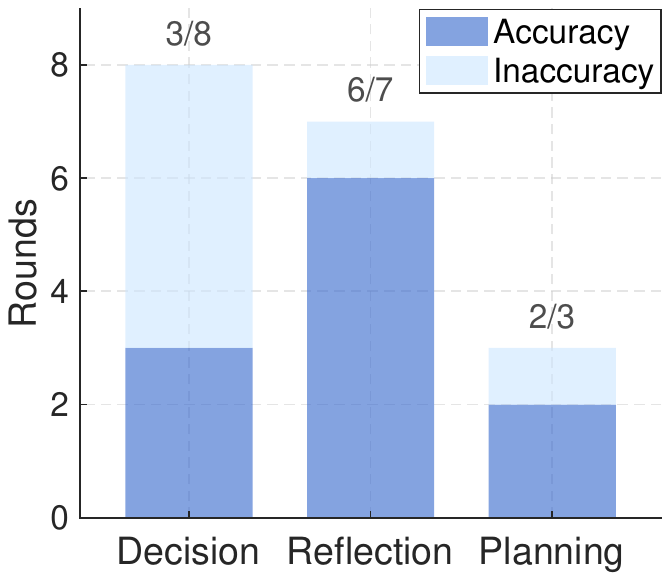}\label{Case_2}}
	\caption{Performance evaluation of mobile AI agents.}\label{Case}
\end{figure}

Algorithm \ref{diffusion_algorithm} illustrates the procedure of the EDMSAC algorithm, where its computational complexity primarily consists of three components: In action sampling, the computational overhead arises from environment interaction and the reverse process, with a complexity of $\mathcal{O}(VZ(1+M|\boldsymbol{\theta}|))$~\cite{10409284}. In dynamic pruning, the computational overhead is due to the calculations of neuron importance and masking, leading to a complexity of $\mathcal{O}(\sum_{h=1}^H(x^{(h)}y^{(h)}+x^{(h)}\log x^{(h)}))$~\cite{10416899}. In parameter updates, the computational complexity stems from updating network parameters, given by $\mathcal{O}(V(B+1)(|\boldsymbol{\theta}| + |\boldsymbol{\upsilon}|))$~\cite{10409284}. Therefore, the computational complexity of the EDMSAC algorithm is $\mathcal{O}(VZ(1+M|\boldsymbol{\theta}|) + \sum_{h=1}^H(x^{(h)}y^{(h)}+x^{(h)}\log x^{(h)}) + V(B+1)(|\boldsymbol{\theta}| + |\boldsymbol{\upsilon}|))$.

\section{Simulation Results}\label{Simulation}
In this section, we first evaluate the performance of mobile AI agents, where we leverage LLMs called Qwen-VL to construct an AI agent capable of operating a mobile phone to execute various tasks. We then compare the proposed EDMSAC algorithm with several DRL algorithms in optimal dynamic contract design. Finally, we validate the effectiveness of the proposed two-period dynamic contract scheme.

\subsection{Experimental Setup}
\textit{Experiment Settings:} We consider ESs with homogeneous hardware capabilities and classify them into two types: high WTP and low WTP~\cite{JinboDiffIoT, wen2024sustainablediffusionbasedincentivemechanism}. Specifically, $\theta^t_1$ and $\theta^t_2$ are randomly sampled within $[15, 20)$ and $[20, 25)$, respectively~\cite{10517486}. $p^1_k$ and $p^2_k$ are randomly generated following the Dirichlet distribution~\cite{10517486}. The setting of simulation parameters is summarized in Table \ref{parameter}. For the hyperparameter setting of the EDMSAC algorithm, DM-based actor networks adopt the variance proportional noise schedule strategy~\cite{10409284}, where the noise schedule strategy determines the amount of noise injected into input data at each timestep during the forward diffusion process~\cite{wen2024sustainablediffusionbasedincentivemechanism}. Both the actor and critic networks are trained using the Adam optimizer, with a weight decay rate of $0.0001$ to regularize model weights~\cite{10409284}. The learning rates are set to $2\times 10^{-7}$ for the actor networks and $2\times 10^{-6}$ for the critic networks, respectively~\cite{10409284}. In addition, the discount factor $\gamma$ is set to $0.99$~\cite{10529221}, the update rate of the target networks $\tau$ is set to $0.005$~\cite{10529221}, the temperature coefficient $\kappa$ is set to $0.05$~\cite{10529221}, the denoising step $M$ is set to $6$~\cite{10409284}, and the pruning rate $\varrho$ is set to $0.5$~\cite{wen2024sustainablediffusionbasedincentivemechanism}. Our experiments use PyTorch with CUDA 12.0 on an NVIDIA GeForce RTX 3080 Laptop GPU.

\begin{table}[t]\label{parameter}
	\renewcommand{\arraystretch}{1.2}
	\caption{Simulation Parameters}
    \centering
	\begin{tabular}{m{5.7cm}<{\raggedright}|m{2.1cm}<{\centering}}
    \toprule[1pt]
    \rowcolor{gray!6}
		\textbf{Parameters} & \textbf{Values}\\	
		\hline
        Number of ESs ($N$)~\cite{10086671}  &  $\{3, 6, 12, 18\}$\\	
        \hline
		Pre-defined hyperparameters ($\delta, L, \varepsilon, \lambda$)~\cite{chen2025federated}  &  $\{0.02,8,2,0.01\}$\\	
		\hline
		  Constant related to the hardware architecture of ES $n$ ($\varsigma_n$)~\cite{9729765}  &  $10^{-23}$  \\	
		\hline
		Number of CPU cycles required for perceiving feature data ($\Upsilon_n^{Per}$)~\cite{chen2025federated}   & $100\:\rm{cycles/bit}$ \\
		\hline
		  CPU speed of ES $n$ ($f_n$)~\cite{chen2025federated} &  $64\:\rm{MHz}$\\	
		\hline		
		Size of feature data perceived by ES $n$ ($D_n^{Fea}$)~\cite{10829636} &  $1\:\rm{MB}$\\
		\hline
        Effective switched capacitance of ES $n$ ($\rho_n$)~\cite{chen2025federated} &  $10^{-16}$\\
		\hline
        Number of CPU cycles required for creating one bit of agent modules ($\Upsilon_n^{Cre}$)~\cite{chen2025federated} &  $120\:\rm{cycles/bit}$\\ \hline
        Transmit power of ES $n$ ($P_n$)~\cite{10086671} &  $[20, 33]\:\rm{dBm}$\\	
		\hline
		  Model size of agent modules constructed by ES $n$ ($D_n^{Agent}$)~\cite{10829636}  &  $10\:\rm{MB}$  \\	
		\hline
		Transmission rate from ES $n$ to the cloud server through the fiber link ($r_{n, c}$)~\cite{chen2025federated}   & $[1,3]\:\rm{Mbps}$  \\
		\hline
		  Unit cost of energy consumption ($\sigma$)~\cite{DynamicLim} &  $[0.5,1]$\\	
		\hline		
		  Scaling factor that affects the overall magnitude of economic benefits ($\alpha$)~\cite{DynamicLim} &  $\{200, 250\}$\\
		\hline
        Energy consumption of global AI agent integration ($E_{C}$)~\cite{chen2025federated} &  $[20, 25]\:\rm{Joules}$\\
		\hline
        Discount factor ($\beta$)~\cite{DCTITS} &  $\{0, 0.5, 1\}$\\
    \bottomrule[1pt]
	\end{tabular}\label{parameter}
\end{table}

\textit{Algorithm Comparisons:} To evaluate the performance of the proposed EDMSAC algorithm, we compare it with mainstream DRL algorithms in optimal contract design, and present the core hyperparameter settings used for each algorithm:
\begin{itemize}
    \item \textit{DMSAC~\cite{10409284}:} The DMSAC algorithm does not incorporate the pruning rate. All hyperparameters are kept consistent with those used in the EDMSAC algorithm. It is worth noting that the computational complexity of the DMSAC algorithm is $\mathcal{O}(VZ(1+M|\boldsymbol{\theta}|) + V(B+1)(|\boldsymbol{\theta}| + |\boldsymbol{\upsilon}|))$.
    \item \textit{DMDDPG~\cite{10529221}:} We set the exploration noise to a fixed Gaussian distribution with zero mean and standard deviation of $0.1$. All other basic hyperparameters are the same as those used in the EDMSAC algorithm.
    \item \textit{PPO~\cite{wen2024learningbasedbigdatasharing}:} We apply the $\tanh$ activation function as the action bounding method, and set the discount factor $\gamma$ and the learning rate to $0.95$ and $3\times10^{-5}$, respectively. All other basic hyperparameters are default, which are commonly used in prior literature~\cite{wen2024learningbasedbigdatasharing}.
    \item \textit{SAC~\cite{JinboDiffIoT}:} We set the learning rate for the temperature coefficient $\kappa$ to $10^{-4}$. The learning rates are set to $10^{-4}$ for the actor networks and $10^{-3}$ for the critic networks, respectively. All other basic hyperparameters are the same as those used in the EDMSAC algorithm.
    \item \textit{DDPG~\cite{wen2024sustainablediffusionbasedincentivemechanism}:} We set the learning rates to $10^{-4}$ for the actor networks and $10^{-3}$ for the critic networks, respectively. The discount factor $\gamma$ is set to $0.99$. All other basic hyperparameters are the same as those used in the DMDDPG algorithm.
\end{itemize}

\textit{Scheme Comparisons:} To evaluate the performance of the proposed dynamic contract scheme, we compare it with other typical contract schemes:
\begin{itemize}
    \item \textit{Two-period static contract scheme~\cite{DCTITS}:} This scheme involves executing the static contract twice, and we set $\beta = 0$ to simplify it to a one-stage static contract.
    \item \textit{Two-period random contract scheme~\cite{JinboDiffIoT}:} This scheme randomly generates two-period contracts for ESs. When $\beta \neq 0$, this scheme operates dynamically across two periods. Conversely, when $\beta = 0$, this scheme is static.
\end{itemize}

\begin{figure*}[t]
		\centering
        \vspace{-0.5cm}
		\begin{subfigure}[EDMSAC vs PPO, SAC, and DDPG.]{\includegraphics[width=0.305\linewidth]{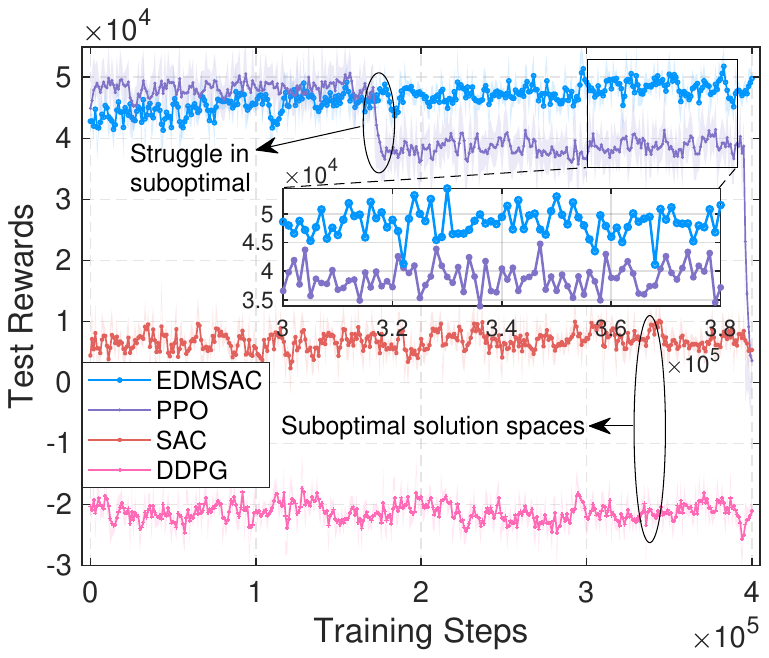}\label{DRL_algorithm}}
		\end{subfigure}
		\hfill
		\begin{subfigure}[EDMSAC vs DMSAC and DMDDPG.]{\includegraphics[width=0.31\linewidth]{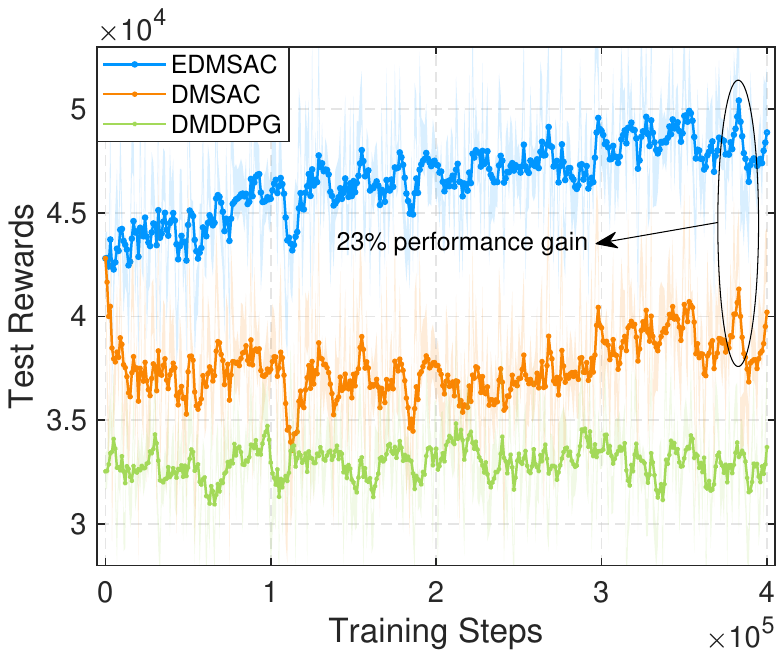}\label{Diffusion_policy}}
		\end{subfigure}
		\hfill
		\begin{subfigure}[Performance results.]{\includegraphics[width=0.312\linewidth]{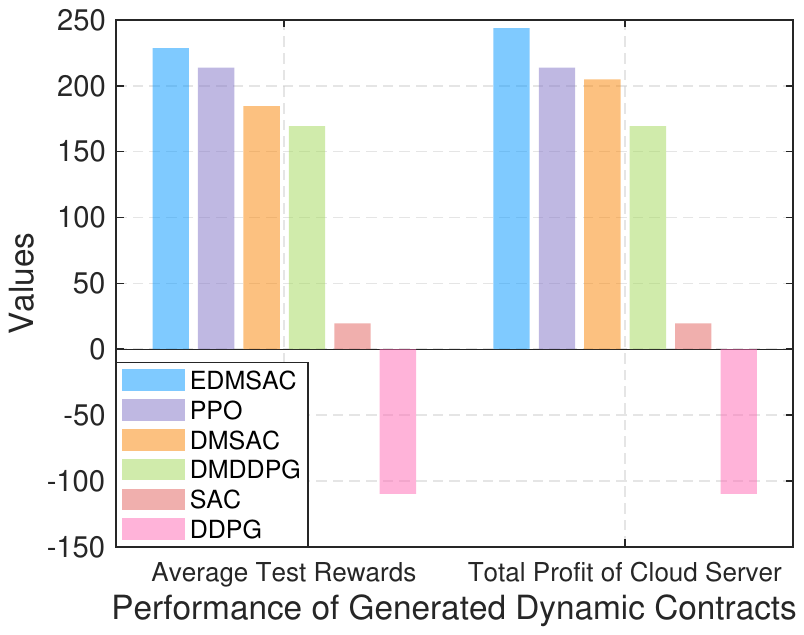}\label{Value}}
		\end{subfigure}
		\caption{Performance comparisons of the EDMSAC algorithm with several DRL algorithms in optimal dynamic contract design.}
		\label{Performance of EDMSAC}
\end{figure*}

\begin{figure}[t]
    \centering  
    \includegraphics[width=0.3\textwidth]{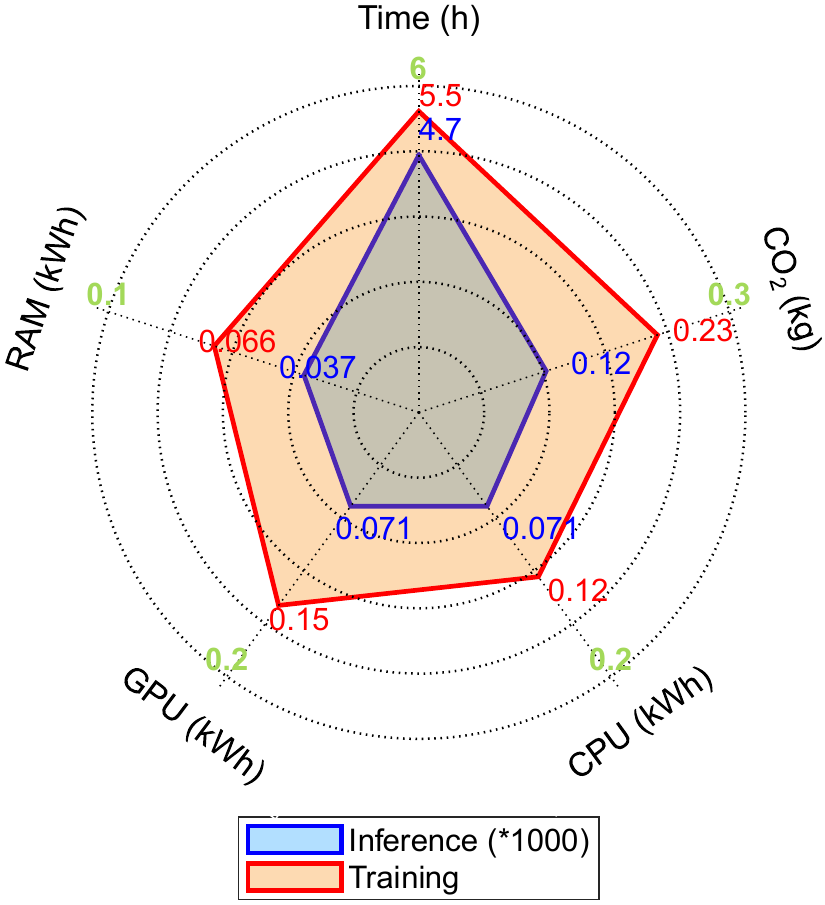}   
    \caption{The resource consumption of the proposed EDMSAC algorithm for performing one training and multiple inferences.}
    \label{radar}
\end{figure}
\subsection{Practical Implementation Guidelines}
We present a practical implementation guideline for the proposed federated AI agent construction framework, investigating how the proposed framework can be closely integrated with the proposed dynamic contract scheme and the EDMSAC algorithm in practical scenarios.

\begin{itemize}
    \item \textbf{Infrastructure Initialization:} Each ES is equipped with sufficient computing resources (e.g., embedded GPUs) and preloaded with basic AI models from the cloud server. Upon boot-up, ESs establish low-latency communication with the cloud server through fiber links.
    \item \textbf{Dynamic Contract Generation:} Based on the predicted WTP levels of ESs, the cloud server executes the proposed EDMSAC algorithm to generate the optimal two-period dynamic contract. Then, the cloud server releases the generated dynamic contract to ESs at the beginning of federated AI agent construction.
    \item \textbf{Agent Module Construction:} The ESs first select their preferred contract item according to their WTP levels. Following contract confirmation, the ESs construct agent modules using perceived feature data. Then, the ESs upload constructed agent modules to the cloud server. The process can be referred to Section \ref{framework_design}.
    \item \textbf{User Interaction Interfaces:} The cloud server integrates uploaded modules into complete AI agents and deploys them to ESs. Users can interact with them through intuitive mobile-based interfaces, such as a customized application connected via local IP networks, enabling real-time AI agent services, including personal assistants.
\end{itemize}

\subsection{Robustness and Security Analysis}
Due to its contract-theoretic design and EDMSAC-based optimization, the proposed scheme provides intrinsic robustness and security properties:
\begin{enumerate}[1)]
\item \textit{The IC and IIC constraints of the proposed scheme inherently mitigate strategic misreporting and free-riding behaviors.} Since ESs maximize their long-term utilities only by truthfully revealing their WTP levels, any deviation or misreporting leads to a strictly lower expected utility.
\item\textit{The contract realization process across two periods enhances robustness against opportunistic behaviors.} The cloud server issues contract-based rewards to the ESs only after successful agent module integration in each period, which discourages non-cooperative actions such as incomplete training.
\item \textit{Uncertainties arising from environments are reflected in the environmental state.} The EDMSAC algorithm learns a contract policy that maximizes expected long-term utility, which exhibits robustness to moderate environmental perturbations without requiring explicit failure modeling.
\end{enumerate}

\subsection{Performance Evaluation of Mobile AI Agents}
As illustrated in Fig. \ref{case_study}, we present two case studies in which a mobile AI agent operates a real smartphone to perform tasks. This experiment involves the distributed execution of agent modules, as well as task decomposition and coordination during execution. For the perception module, we leverage Qwen-VL-Plus to create it, and we use the ConvNextViT-document model and the GroundingDINO model as text and icon recognition tools, respectively~\cite{wang2024mobileagentv2mobiledeviceoperation}. For the decision-making, reflection, and planning modules, we leverage Qwen-VL-Max to create them. To evaluate the performance of the constructed AI agent, we adopt several evaluation metrics: Decision Accuracy (DA), which measures the accuracy of the decision-making module; Reflection Accuracy (RA), which assesses the accuracy of the reflection module; and Planning Accuracy (PA), which evaluates the accuracy of the planning module. As shown in Fig. \ref{Case}, the performance of the AI agent in executing simple tasks, such as Case 1, is superior to that in handling more complex tasks, as in Case 2, as evidenced by higher values of DA, RA, and PA. Notably, when executing simple tasks, the AI agent may perform additional rounds to verify task completion, leading to increased service latency.

\subsection{Performance Evaluation of the EDMSAC Algorithm}
In Fig. \ref{Performance of EDMSAC}, we present a performance comparison between the proposed EDMSAC algorithm and several representative DRL algorithms in optimal contract design, including DMSAC, DMDDPG, PPO, SAC, and DDPG. From Figs. \ref{DRL_algorithm} and \ref{Diffusion_policy}, we observe that the proposed EDMSAC algorithm outperforms these DRL algorithms in optimal dynamic contract design because DRL algorithms struggle in different suboptimal solution spaces, especially PPO algorithms. The superior performance of the proposed EDMSAC algorithm can be primarily attributed to two key reasons. First, the EDMSAC algorithm has a strong capability to capture complex environmental information because of GDMs~\cite{10529221}, which helps avoid the identification of suboptimal dynamic contracts. Moreover, its fine-grained policy tuning mitigates the impact of environmental randomness, enhancing decision robustness. Second, through structural pruning techniques, the unimportant neurons in the DM-based actor networks are masked rather than permanently removed, thereby preserving the network structure while reducing computational redundancy. Therefore, the EDMSAC algorithm can achieve higher denoising efficiency, enabling the generation of more optimal and stable dynamic contracts. From Fig. \ref{Value}, we also observe that the EDMSAC algorithm achieves a higher total profit of the cloud server compared with other DRL algorithms. Overall, the above analysis demonstrates the effectiveness and superiority of the proposed EDMSAC algorithm in optimizing dynamic contract design, which enhances the system performance of federated AI agent construction under dynamic information asymmetry. 

In Fig. \ref{radar}, we employ the Python package named CodeCarbon\footnote{\url{https://github.com/mlco2/codecarbon}} to estimate the resource consumption of the proposed EDMSAC algorithm for one training session and $20$ inference runs. The evaluation considers five dimensions: time ($\rm{h}$), $\text{CO}_{\text{2}}$ ($\rm{kg}$), CPU ($\rm{kWh}$), GPU ($\rm{kWh}$), and random access memory ($\rm{kWh}$). To enhance clarity, we scale the inference results of all five metrics by a factor of $1000$. We observe that all metrics for one training session are significantly higher than those for $20$ inference sessions. Especially, the total carbon emissions for performing $20$ inferences are only $0.123\:\rm{g}$. The reason is that the structural pruning technique can improve computational efficiency of the denoising process, thereby reducing total carbon emissions. These findings indicate the environmental sustainability of the proposed EDMSAC algorithm in optimal dynamic contract design.

\begin{figure*}[t]
		\centering
		\begin{subfigure}[Pruning rates.]{\includegraphics[width=0.312\linewidth]{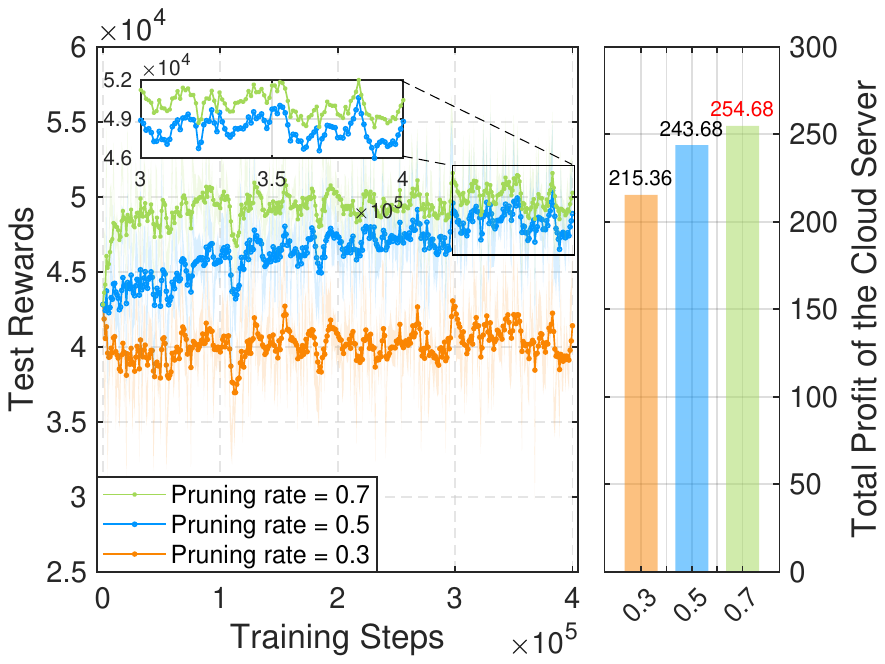}\label{Pruning_rate}}
		\end{subfigure}
		\hfill
		\begin{subfigure}[Denoising steps.]{\includegraphics[width=0.312\linewidth]{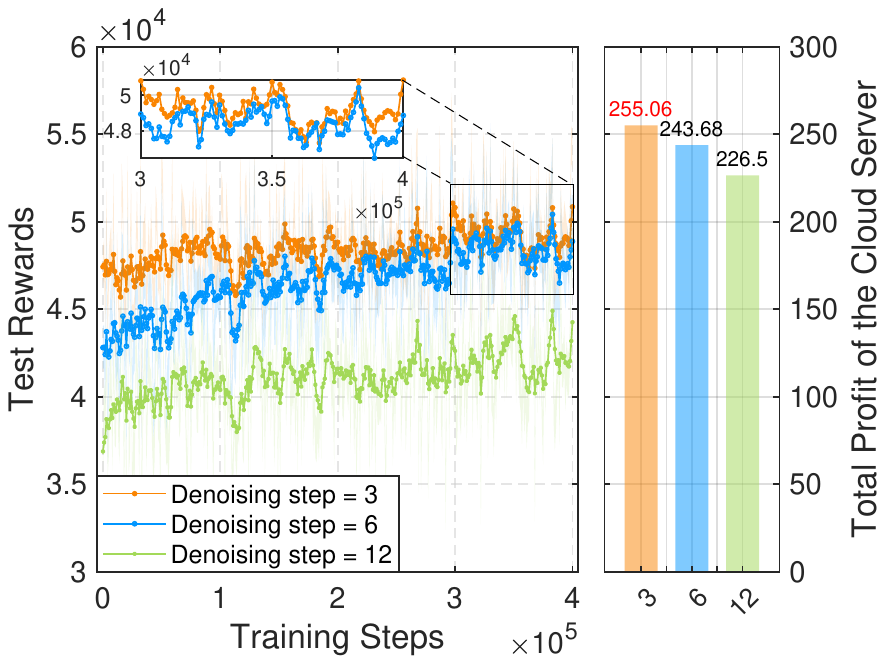}\label{diffusion_denoising}}
		\end{subfigure}
		\hfill
		\begin{subfigure}[Learning rates.]{\includegraphics[width=0.312\linewidth]{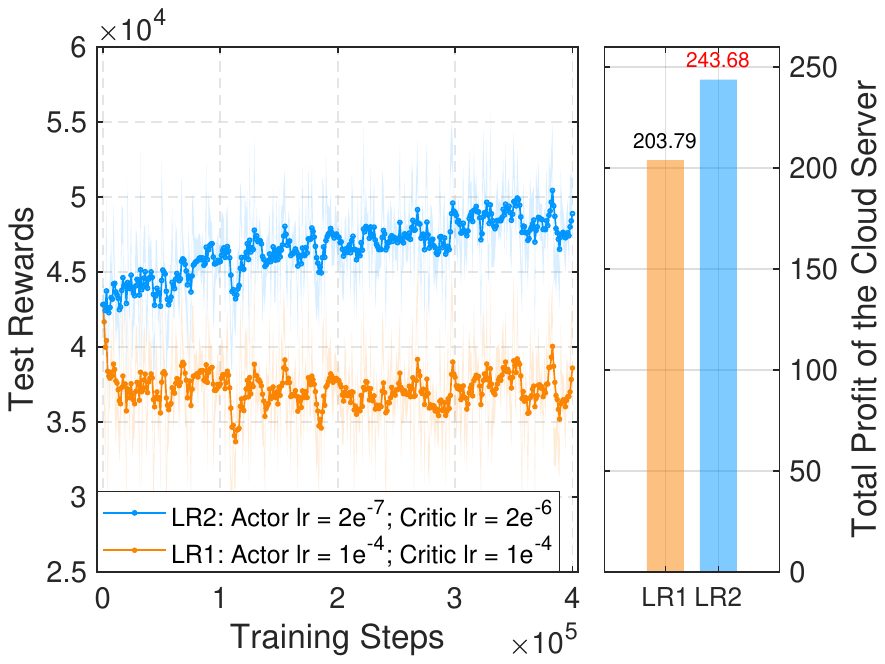}\label{Learning_rate}}
		\end{subfigure}
		\caption{Performance of the EDMSAC algorithm in optimal dynamic contract design under critical hyperparameter settings, where the optimal value corresponding to the optimal hyperparameter is marked in \textcolor{red}{red}.}
		\label{diffusion_configuration}
\end{figure*}

As shown in Fig. \ref{diffusion_configuration}, we present the performance of the EDMSAC algorithm in optimal dynamic contract design with different hyperparameter configurations. Figure \ref{Pruning_rate} illustrates the performance of the EDMSAC algorithm under different pruning rates, i.e., $30\%$, $50\%$, and $70\%$. We observe that the EDMSAC algorithm demonstrates superior performance under higher pruning rates, leading to faster convergence and higher test rewards. The reason is that a higher pruning rate indicates that more redundant and unimportant neurons in DM-based actor networks are structurally masked. As a result, the pruned DM-based actor networks exhibit better generalization capabilities and are more effective at learning essential features in optimal dynamic contract design. Figure \ref{diffusion_denoising} presents the performance of the EDMSAC algorithm under different denoising steps, i.e., $3$, $6$, and $12$. We observe that the EDMSAC algorithm achieves faster convergence and yields higher total profits for the cloud server under a smaller denoising step. The reason is that a smaller denoising step can achieve higher computational efficiency of DM-based actor networks. Figure \ref{Learning_rate} illustrates the performance of the EDMSAC algorithm under two groups of different learning rates. The first group is the actor networks with a learning rate of $2\times10^{-7}$ and the critic networks with a learning rate of $2\times10^{-6}$. The second group is the actor networks and critic networks with the same learning rate of $10^{-4}$. We observe that the EDMSAC algorithm has better performance under the first group of learning rates, indicating that the pruned actor networks need to learn an efficient policy in optimal dynamic contract design under small learning rates. 

\subsection{Performance Evaluation of the Dynamic Contract Scheme}
\begin{figure*}[t]
		\centering
        \vspace{-0.5cm}
		\begin{subfigure}[State 1.]{\includegraphics[width=0.303\linewidth]{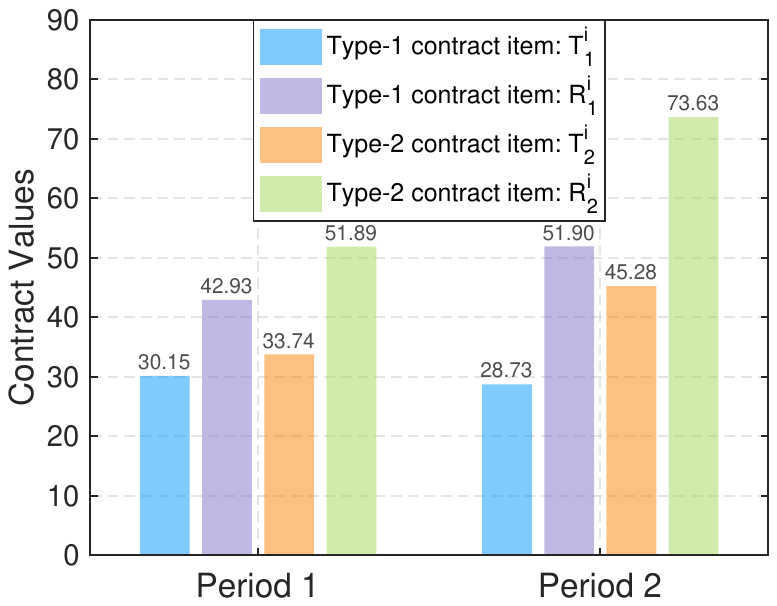}\label{state1}}
		\end{subfigure}
		\hfill
		\begin{subfigure}[State 2.]{\includegraphics[width=0.31\linewidth]{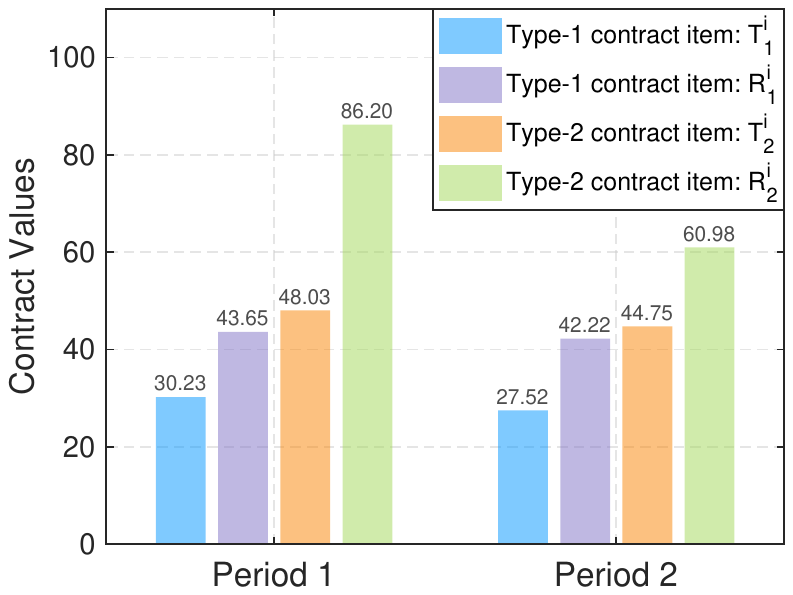}\label{state2}}
		\end{subfigure}
		\hfill
		\begin{subfigure}[State 3.]{\includegraphics[width=0.31\linewidth]{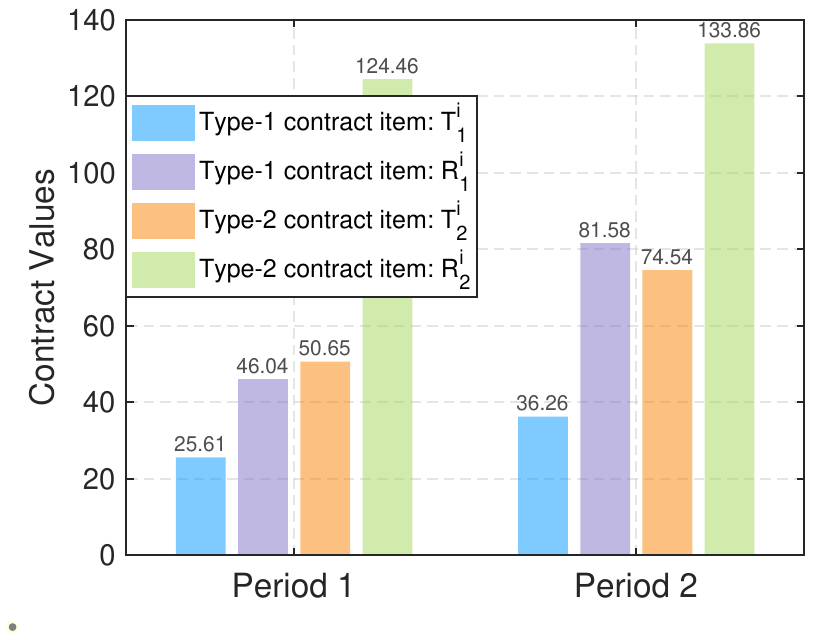}\label{State3}}
		\end{subfigure}
		\caption{Two-period dynamic contracts generated by the EDMSAC algorithm under three network states.}
		\label{Generated_contract}
\end{figure*}

In Fig. \ref{Generated_contract}, we illustrate the two-period dynamic contracts generated by the EDMSAC algorithm under three network states, with a pruning rate of $0.5$ and a denoising step of $6$. Additionally, the learning rates of the actor and critic networks are set to $2\times10^{-7}$ and  $2\times10^{-6}$, respectively. Attributed to the exploration and generalization capabilities of GDMs, the proposed EDMSAC algorithm demonstrates the capability to generate optimal dynamic contracts across varying network states, indicating that our algorithm possesses strong adaptability and generalization abilities to the dynamic environments inherent in federated AI agent construction, thereby ensuring robust system performance in mobile metaverses. We observe that in period 2, the ESs with higher types perform more training rounds $T_2^k$ for agent module creation and consequently receive higher rewards $R_2^k$, which validates Lemmas \ref{lemma1} and \ref{lemma2}, respectively. In period 1, the ESs with higher types also perform more training rounds $T_i^k$ for agent module creation and receive higher rewards $R_i^k$, which demonstrates Lemmas \ref{lemma7} and \ref{lemma8}, respectively.

\begin{figure*}[t]
		\centering
		\begin{subfigure}[Test reward comparison between the proposed scheme and three baseline contract schemes.]{\includegraphics[width=0.31\linewidth]{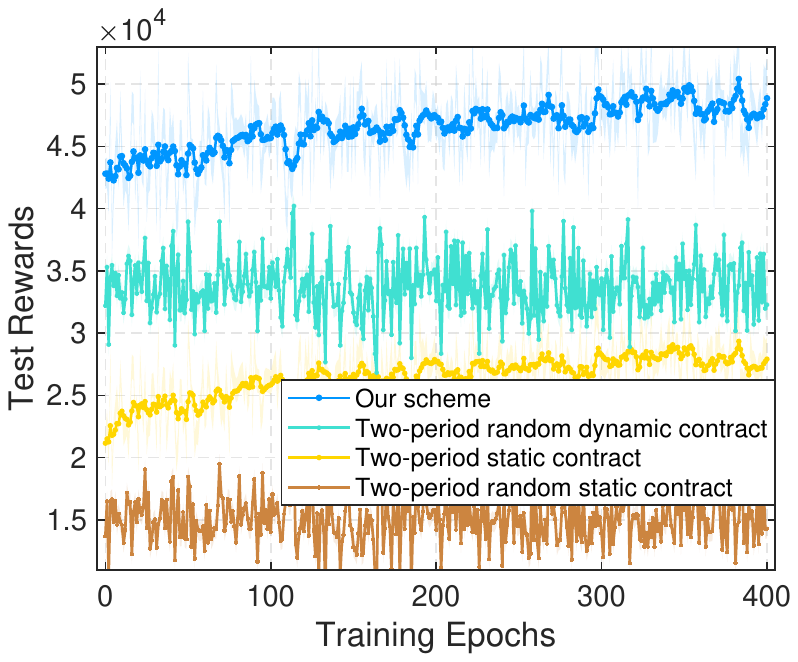}\label{Compare_scheme}}
		\end{subfigure}
		\hfill
		\begin{subfigure}[Relationship between the total profit of the cloud server and the number of ESs.]{\includegraphics[width=0.313\linewidth]{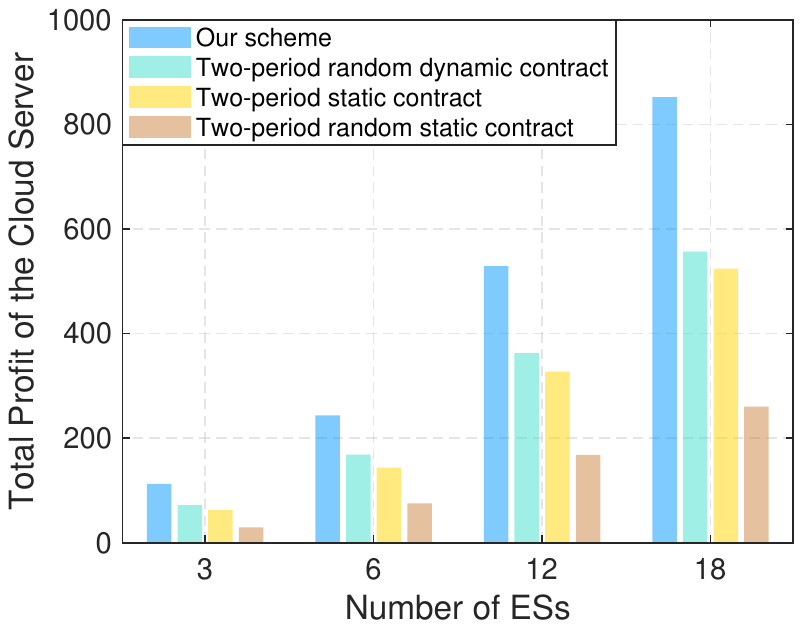}\label{number_ES}}
		\end{subfigure}
		\hfill
		\begin{subfigure}[Impacts of $\alpha$ and $\beta$ on the total profit of the cloud server under different numbers of ESs.]{\includegraphics[width=0.316\linewidth]{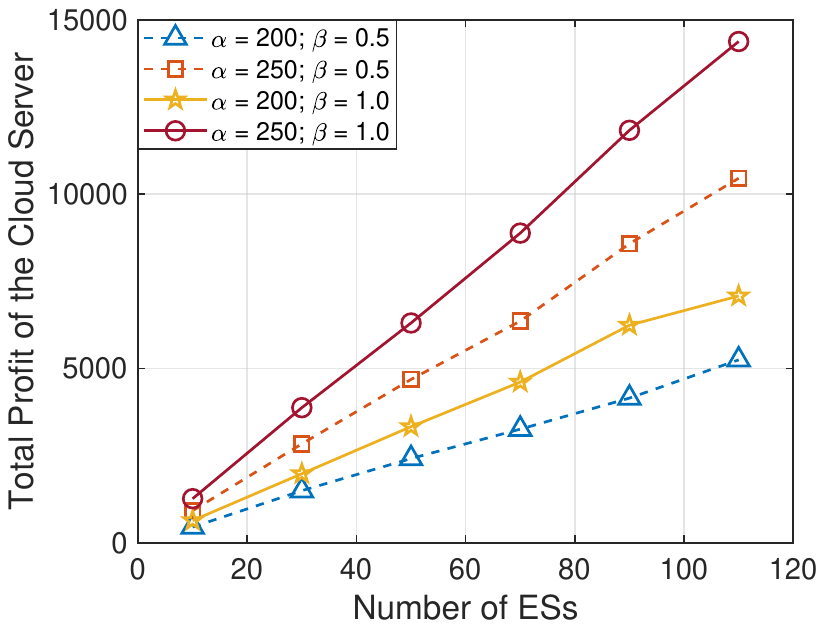}\label{alpha_beta}}
		\end{subfigure}
		\caption{Performance evaluation of the proposed two-period dynamic contract scheme.}
		\label{Scheme_performance}
\end{figure*}

In Fig. \ref{Scheme_performance}, we compare the proposed two-period dynamic contract scheme with three baseline contract schemes. Figure \ref{Compare_scheme} presents the test reward comparison between the proposed scheme and three baseline contract schemes. Our dynamic contract scheme consistently achieves higher test rewards compared with the other three contract schemes, indicating that the proposed two-period dynamic contract scheme possesses superior performance. In addition, the test rewards obtained by dynamic contract schemes consistently surpass those of static contract schemes, highlighting the advantages of incorporating adaptability in contract design. The reason is that, unlike static contract schemes, dynamic contract schemes take into account the profit of the cloud server in the second period. As a result, the total profit of the cloud server achieved under the dynamic contract scheme is higher than that of the static contract scheme. Figure \ref{number_ES} illustrates the relationship between the total profit of the cloud server and varying numbers of ESs under four contract schemes. No matter which contract scheme, as the number of ESs increases, the total profit of the cloud server also increases. Moreover, compared with the three baseline contract schemes, the proposed two-period dynamic contract scheme can achieve the highest total profit for the cloud server. This is primarily because the proposed scheme is capable of designing suitable contracts for ESs based on their respective types rather than generating contracts randomly, which enables more effective resource utilization for federated AI agent construction and contributes to higher total profits for the cloud server. Figure \ref{alpha_beta} presents the impacts of the scaling factor $\alpha$ and the discount factor $\beta$ on the total profit of the cloud server under different numbers of ESs. When $\beta$ remains constant, the total profit of the cloud server increases with the rise of $\alpha$. The reason is that a larger $\alpha$ reflects a greater overall magnitude of economic benefits allocated to the cloud server, thereby enhancing its total profit. In addition, when $\alpha$ remains constant, the total profit of the cloud server increases as $\beta$ increases. This is primarily because the cloud server obtains higher profits in the second period of federated AI agent construction, thus enhancing its total profit.

\section{Conclusion and Future Work}\label{Conclusion}
In this paper, we have studied federated AI agent construction in mobile metaverses. Specifically, we have proposed an edge-cloud collaboration-based federated AI agent construction framework in mobile metaverses. In this framework, ESs serve as agent infrastructures to construct agent modules in a distributed manner, and the cloud server subsequently integrates them into complete AI agents and deploys the constructed AI agents at the edge, thereby providing AI agent services for users in mobile metaverses. Considering the dynamic information asymmetry in federated AI agent construction, we have designed a two-period dynamic contract model to incentivize ESs to continuously participate in federated AI agent construction. Furthermore, to effectively identify optimal dynamic contracts, we have proposed the EDMSAC algorithm, in which the redundant neurons of DM-based actor networks are dynamically pruned, thereby enhancing denoising efficiency for dynamic contract sampling. Extensive simulations have demonstrated that the proposed EDMSAC can achieve up to a $23\%$ improvement over the DMSAC algorithm in optimal dynamic contract generation. For future work, we aim to develop a mobile AI agent paradigm built upon the proposed framework, which will empower AI agents with enhanced adaptability and decision-making capabilities in real-world mobile environments. Furthermore, we plan to build a working prototype using real edge devices to transform the proposed framework into a deployable system.

\bibliographystyle{IEEEtran}
\bibliography{ref}

\newpage
\clearpage
\newpage
\appendices
\section{Proof for Lemma \ref{lemma1}}\label{Proof_for_lemma1}
\renewcommand{\theequation}{A-\arabic{equation}}
\setcounter{equation}{0}
We first prove that if $T_i^2(\theta_k^1) \geq T_j^2(\theta_k^1)$, then $R_i^2(\theta_k^1) \geq R_j^2(\theta_k^1)$. Based on the IC constraints in (\ref{IC2}), we have
    \begin{equation}
        \theta_i^2 R_i^2(\theta_k^1) - c T_i^2(\theta_k^1) \geq \theta_i^2 R_j^2(\theta_k^1) - c T_j^2(\theta_k^1).
    \end{equation}
    We transform the above inequality into
    \begin{equation}
        \theta_i^2 [R_i^2(\theta_k^1)-R_j^2(\theta_k^1)] \geq c[T_i^2(\theta_k^1)-T_j^2(\theta_k^1)].
    \end{equation}
    Since $T_i^2(\theta_k^1) \geq T_j^2(\theta_k^1)$, we have $\theta_i^2 [R_i^2(\theta_k^1)-R_j^2(\theta_k^1)] \geq 0$, then $R_i^2(\theta_k^1) \geq R_j^2(\theta_k^1)$. Similarly, given that $R_i^2(\theta_k^1) \geq R_j^2(\theta_k^1)$, we have $T_i^2(\theta_k^1) \geq T_j^2(\theta_k^1)$. Thus, the proof of Lemma \ref{lemma1} is completed.

\section{Proof for Lemma \ref{lemma2}}\label{Proof_for_lemma2}
\renewcommand{\theequation}{B-\arabic{equation}}
\setcounter{equation}{0}
Based on the IC constraints in (\ref{IC2}), we have
    \begin{equation}\label{lemm2_1}
        \theta_i^2 R_i^2(\theta_k^1) - c T_i^2(\theta_k^1) \geq \theta_i^2 R_j^2(\theta_k^1) - c T_j^2(\theta_k^1),
    \end{equation}
    \begin{equation}\label{lemm2_2}
        \theta_j^2 R_j^2(\theta_k^1) - c T_j^2(\theta_k^1) \geq \theta_j^2 R_i^2(\theta_k^1) - c T_i^2(\theta_k^1).
    \end{equation}
    By combining (\ref{lemm2_1}) and (\ref{lemm2_2}), we have
    \begin{equation}\label{lemm2_3}
        (\theta_i^2 - \theta_j^2)(R_i^2(\theta_k^1) - R_j^2(\theta_k^1)) \geq 0.
    \end{equation}
    Since $\theta_i^2(\theta_k^1) \geq \theta_j^2(\theta_k^1)$, we have $R_i^2(\theta_k^1) \geq R_j^2(\theta_k^1)$. Thus, the proof of Lemma \ref{lemma2} is completed.

\section{Proof for Lemma \ref{lemma4}}\label{Proof_for_lemma4}
\renewcommand{\theequation}{C-\arabic{equation}}
\setcounter{equation}{0}
 Given the IC constraints in (\ref{IC2}) and the ascending order of types of ESs, we have
        \begin{equation}\label{lemm4_2}
        \begin{split}
        \theta_i^2 R_i^2(\theta_k^1) - c T_i^2(\theta_k^1) &- E \geq \theta_i^2 R_1^2(\theta_k^1) - c T_1^2(\theta_k^1) - E\\
        &\geq \theta_1^2 R_1^2(\theta_k^1) - c T_1^2(\theta_k^1) - E \geq 0.
        \end{split}
        \end{equation}
        
        From (\ref{lemm4_2}), we can observe that if the type-$1$ ES with the lowest WTP has a non-negative utility, it implies that the other IR constraints automatically hold. Additionally, for $\theta_1^2 R_1^2(\theta_k^1) - c T_1^2(\theta_k^1) - E \geq 0$, the cloud server would reduce the reward $R_1^2(\theta_k^1)$ as much as possible until $\theta_1^2 R_1^2(\theta_k^1) - c T_1^2(\theta_k^1) - E = 0$ ~\cite{JinboDiffIoT}, thus maximizing its utility. Thus, the proof of Lemma \ref{lemma4} is completed.

\section{Proof for Lemma \ref{lemma5}}\label{Proof_for_lemma5}
\renewcommand{\theequation}{D-\arabic{equation}}
\setcounter{equation}{0}
For the proof of LDIC, we can first obtain the following equations based on the IC constraints in (\ref{IC2}):
        \begin{equation}\label{lemma5_1}
        \theta_i^2 R_i^2(\theta_k^1) - c T_i^2(\theta_k^1) \geq \theta_i^2 R_{i-1}^2(\theta_k^1) - c T_{i-1}^2(\theta_k^1).
        \end{equation}
        \begin{equation}\label{lemma5_2}
        \theta_{i-1}^2 R_{i-1}^2(\theta_k^1) - c T_{i-1}^2(\theta_k^1) \geq \theta_{i-1}^2 R_{i-2}^2(\theta_k^1) - c T_{i-2}^2(\theta_k^1),
        \end{equation}
        where (\ref{lemma5_2}) can be transformed into
        \begin{equation}
        \theta_{i-1}^2 (R_{i-1}^2(\theta_k^1)-R_{i-2}^2(\theta_k^1)) \geq c(T_{i-1}^2(\theta_k^1)-T_{i-2}^2(\theta_k^1)).
        \end{equation}
        Based on the ascending order of types, we can further obtain
        \begin{equation}\label{lemma5_4}
        \theta_{i}^2 (R_{i-1}^2(\theta_k^1)-R_{i-2}^2(\theta_k^1)) \geq c(T_{i-1}^2(\theta_k^1)-T_{i-2}^2(\theta_k^1)),
        \end{equation}
        which can be transformed into
        \begin{equation}
        \theta_{i}^2 R_{i-1}^2(\theta_k^1) - cT_{i-1}^2(\theta_k^1) \geq \theta_{i}^2 R_{i-2}^2(\theta_k^1) - cT_{i-2}^2(\theta_k^1).
        \end{equation}
        By repeating the above process for type $j \in \{i-2, i-3, \ldots,1\}$, we can obtain the global DIC as follows:
        \begin{equation}
        \begin{split}
         \theta_{i}^2 R_i^2(\theta_k^1) - c T_i^2(\theta_k^1) \geq &\theta_{i}^2 R_{i-1}^2(\theta_k^1) - c T_{i-1}^2(\theta_k^1) \geq \\
         &\cdots \geq \theta_{i}^2 R_1^2(\theta_k^1) - c T_1^2(\theta_k^1).
         \end{split}
        \end{equation}

        A similar process can be taken to prove that if the LUIC constraints hold, the global UIC constraints also hold, i.e.,
        \begin{equation}
         \begin{split}
         \theta_{i}^2 R_i^2(\theta_k^1) - c T_i^2(\theta_k^1) \geq &\theta_{i}^2 R_{i+1}^2(\theta_k^1) - c T_{i+1}^2(\theta_k^1) \geq \\
         &\cdots \geq \theta_{i}^2 R_K^2(\theta_k^1) - c T_K^2(\theta_k^1).
         \end{split} 
        \end{equation}
        Thus, the proof of Lemma \ref{lemma5} is completed.

\section{Proof for Lemma \ref{lemma6}}\label{Proof_for_lemma6}
\renewcommand{\theequation}{E-\arabic{equation}}
\setcounter{equation}{0}
 To prove Lemma \ref{lemma6}, we can use the contradiction method to demonstrate that if $\theta_{i}^2 R_i^2(\theta_k^1) - c T_i^2(\theta_k^1) - E =  \theta_{i}^2 R_{i-1}^2(\theta_k^1) - c T_{i-1}^2(\theta_k^1) - E,\:\forall i \in \{2,\ldots, K\}$ and the monotonicity hold, the LUIC constraints hold.

    If $\theta_{i}^2 \geq \theta_{i-1}^2$, then $R_i^2(\theta_k^1) \geq R_{i-1}^2(\theta_k^1)$, and we have 
    \begin{equation}\label{lemma6_1}
       \theta_{i}^2 (R_i^2(\theta_k^1)-R_{i-1}^2(\theta_k^1)) \geq \theta_{i-1}^2(R_i^2(\theta_k^1)-R_{i-1}^2(\theta_k^1)),
    \end{equation}
    which (\ref{lemma6_1}) can be further transformed into
    \begin{equation}
        c(T_i^2(\theta_k^1) -  T_{i-1}^2(\theta_k^1)) \geq \theta_{i-1}^2(R_i^2(\theta_k^1)-R_{i-1}^2(\theta_k^1)),
    \end{equation}
    and we have
    \begin{equation}
    \theta_{i-1}^2 R_{i-1}^2(\theta_k^1) - cT_{i-1}^2(\theta_k^1) \geq \theta_{i-1}^2 R_{i}^2(\theta_k^1) - cT_{i}^2(\theta_k^1).
    \end{equation}
    Therefore, the LUIC constraints remain valid. The proof of Lemma \ref{lemma6} is completed.

\section{Proof for Theorem \ref{theorem2}}\label{Proof_for_Theorem2}
\renewcommand{\theequation}{F-\arabic{equation}}
\setcounter{equation}{0}
We use the contradiction method to demonstrate the uniqueness of the optimal reward in period 2. We assume that there is another optimal reward $\hat{R}_i^2$ that facilitates the greater profit of the cloud server. Since the utility of the cloud server is inversely proportional to the total reward, we can obtain $\sum_{i=1}^K \hat{R}_i^2 < \sum_{i=1}^K (R_i^2)^{\star}$, indicating that there is at least one type $\theta_i^2$ satisfying $\hat{R}_i^2 < (R_i^2)^{\star},\: i\in \mathcal{K}$~\cite{DynamicLim, DCTITS}.

We consider $\hat{R}_i^2 < (R_i^2)^{\star}$, where $1 \leq i < K$. Based on the LDIC constraints, we have
    \begin{equation}\label{theorem2_1}
           \hat{R}_i^2 \geq \hat{R}_{i-1}^2 + \frac{c(T_{i}^2(\theta_k^1) - T_{i-1}^2(\theta_k^1))}{\theta_i^2}.
    \end{equation}
Based on Lemma \ref{lemma6}, we have
    \begin{equation}\label{theorem2_2}
         (R_i^2)^{\star} - (R_{i-1}^2)^{\star} =  \frac{c(T_{i}^2(\theta_k^1) - T_{i-1}^2(\theta_k^1))}{\theta_i^2}.
    \end{equation}
By substituting (\ref{theorem2_2}) into (\ref{theorem2_1}), we have
    \begin{equation}
           \hat{R}_i^2 -  (R_i^2)^{\star} \geq  \hat{R}_{i-1}^2 - (R_{i-1}^2)^{\star}.
    \end{equation}
Since $\hat{R}_i^2 < (R_i^2)^{\star}$, we have $\hat{R}_{i-1}^2 < (R_{i-1}^2)^{\star}$. By recursion, we can eventually obtain $\hat{R}_{1}^2 < (R_{1}^2)^{\star} = \frac{cT_{1}^2(\theta_k^1) + E}{\theta_1^2}$, which violates the IR constraints. Therefore, there does not exist the reward $\hat{R}_i^2$ that facilitates the greater profit of the cloud server. The proof of Theorem \ref{theorem2} is completed.

\section{Proof for Lemma \ref{lemma7}}\label{Proof_for_Lemma7}
\renewcommand{\theequation}{G-\arabic{equation}}
\setcounter{equation}{0}
 By substituting the optimal reward $(R_i^2)^{\star}$ in period 2 into the IIC constraints, we have
    \begin{equation}\label{lemma7_1}
    \begin{split}
        &u_k^1(T_k^1) + \beta\sum_{j=2}^K p_j^2(T_k^1)\\
        &\qquad \times \Bigg[\frac{\theta_j^2E}{\theta_1^2} + \sum_{l=1}^{j-1}\theta_j^2c\bigg(\frac{T_l^2(\theta_k^1)}{\theta_l^2}-\frac{T_l^2(\theta_k^1)}{\theta_{l+1}^2}\bigg)\Bigg]\\
        &\quad \geq u_k^1(T^1_{k^{\prime}}) + \beta\sum_{j=2}^K p_j^2(T_k^1)\\
        &\qquad \times \Bigg[\frac{\theta_j^2E}{\theta_1^2} + \sum_{l=1}^{j-1}\theta_j^2c\bigg(\frac{T_l^2(\theta^1_{k^{\prime}})}{\theta_l^2}-\frac{T_l^2(\theta^1_{k^{\prime}})}{\theta_{l+1}^2}\bigg)\Bigg].
    \end{split}
    \end{equation}
    For ease of notation, we write
    \begin{equation}
        a = \frac{\theta_j^2E}{\theta_1^2} + \sum_{l=1}^{j-1}\theta_j^2c\bigg(\frac{T_l^2(\theta_k^1)}{\theta_l^2}-\frac{T_l^2(\theta_k^1)}{\theta_{l+1}^2}\bigg),
    \end{equation}
    \begin{equation}
        b = \frac{\theta_j^2E}{\theta_1^2} + \sum_{l=1}^{j-1}\theta_j^2c\bigg(\frac{T_l^2(\theta^1_{k^{\prime}})}{\theta_l^2}-\frac{T_l^2(\theta^1_{k^{\prime}})}{\theta_{l+1}^2}\bigg).
    \end{equation}

    Repeating the above procedure can also obtain the IIC constraint for type-$k^{\prime}$ ESs like (\ref{lemma7_1}). Then, by combining the two IIC constraints, we have
    \begin{equation}\label{lemma7_4}
    \begin{split}
      u_k^1(T_k^1)&-u_k^1(T^1_{k^{\prime}})+u^1_{k^{\prime}}(T^1_{k^{\prime}})-u^1_{k^{\prime}}(T^1_{k}) \\
      &+ \beta\sum_{j=2}^K[p_j^2(T_k^1)-p_j^2(T^1_{k^{\prime}})](a-b) \geq 0.
    \end{split}
    \end{equation}

    To better analyze whether (\ref{lemma7_4}) is true, we separate (\ref{lemma7_4}) into three parts, which are given by
    \begin{equation}\label{lemma7_5}
    \begin{split}
      u_k^1(T_k^1)-u_k^1(T^1_{k^{\prime}})&+u^1_{k^{\prime}}(T^1_{k^{\prime}})-u^1_{k^{\prime}}(T^1_{k})\\ 
      &\quad= (\theta_k^1-\theta^1_{k^{\prime}})(R_k^1-R^1_{k^{\prime}}),
    \end{split}
    \end{equation}
    \begin{equation}\label{lemma7_6}
      p_j^2(T_k^1)-p_j^2(T^1_{k^{\prime}}),
    \end{equation}
    and
    \begin{equation}\label{lemma7_7}
        a-b = \theta_j^2 c\sum_{l=1}^{j-1}\Bigg[\bigg(\frac{1}{\theta_l^2}-\frac{1}{\theta_{l+1}^2}\bigg)(T_l^2(\theta_k^1)-T_l^2(\theta^1_{k^{\prime}}))\Bigg].
    \end{equation}

    Since the types of period 1 and period 2 are positively correlated~\cite{DCTITS, DynamicLim}, we have $ p_j^2(T_k^1) > p_j^2(T^1_{k^{\prime}})$, i.e., (\ref{lemma7_6}) is positive. To make (\ref{lemma7_4}) hold, (\ref{lemma7_5}) and (\ref{lemma7_7}) must be positive. Following Lemma \ref{lemma7}, (\ref{lemma7_5}) will be positive if $T_k^1 > T^1_{k^{\prime}}$, and (\ref{lemma7_7}) will be positive if $T_j^2(T_k^1) > T_j^2(T^1_{k^{\prime}})$. Thus, the proof of Lemma \ref{lemma7} is completed.

\section{Proof for Lemma \ref{lemma8}}\label{Proof_for_Lemma8}
\renewcommand{\theequation}{H-\arabic{equation}}
\setcounter{equation}{0}
 Based on the IIC constraints for the ESs with type $\theta_k^1$, we can obtain
     \begin{equation}\label{lemma8_1}
         u_k^1(T_k^1) + \beta\sum_{j=1}^K p_j^2(T_k^1)[u_j^2(T_j^2(T_k^1))-u_j^2(T_j^2(T^1_{k^{\prime}}))] \geq u_k^1(T^1_{k^{\prime}}).
     \end{equation}
     Since $\beta$ and $p_j^2(T_k^1)$ are small, (\ref{lemma8_1}) can be simplified into
     \begin{equation}\label{lemma8_2}
       u_k^1(T_k^1) \geq u_k^1(T^1_{k^{\prime}}),
     \end{equation}
     which can be transformed into
     \begin{equation}\label{lemma8_3}
         \theta_k^1(R_k^1-R^1_{k^{\prime}}) \geq c(T_k^1 - T^1_{k^{\prime}}).
     \end{equation}
     
     We first prove the sufficiency of Lemma \ref{lemma8}, namely if $T_k^1 > T^1_{k^{\prime}}$, then $R_k^1 > R^1_{k^{\prime}}$. From (\ref{lemma8_3}), we can know that $R_k^1 > R^1_{k^{\prime}}$ since $T_k^1 >  T^1_{k^{\prime}}$. Thus, the sufficiency is proved.

     We then prove the necessity of Lemma \ref{lemma8}, namely if $R_k^1 > R^1_{k^{\prime}}$, then $T_k^1 > T^1_{k^{\prime}}$. Based on the IIC constraint for type-$k^{\prime}$ ESs, we can obtain
     \begin{equation}
         u^1_{k^{\prime}}(T^1_{k^{\prime}}) \geq u^1_{k^{\prime}}(T^1_k),
     \end{equation}
     which can also be transformed into
     \begin{equation}\label{lemma8_5}
         \theta^1_{k^{\prime}}(R^1_{k^{\prime}}-R_k^1) \geq c(T^1_{k^{\prime}}-T^1_k).
     \end{equation}
     Given $R^1_{k^{\prime}} > R_k^1$, the left-hand side of (\ref{lemma8_5}) is negative. Thus, we have $T^1_k-T^1_{k^{\prime}}$. The proof of Lemma \ref{lemma8} is completed.

\vfill

\end{document}